\documentclass[aps,prb,amsmath,amssymb,showpacs,superscriptaddress,12pt]{revtex4-1}

\usepackage{graphicx}
\usepackage{amsmath,amssymb,amsfonts}
\usepackage{textcomp}
\usepackage{hyperref}
\usepackage{gensymb}
\usepackage{verbatim}
\usepackage{amsmath}
\hypersetup{breaklinks=true,colorlinks=true,urlcolor=black}
\usepackage{color}
\DeclareGraphicsExtensions{.jpg,.pdf,.png,.eps}

\begin{document}
\title{Phenomenal magneto-elastoresistance of WTe$_{2}$: strain engineering of electronic and quantum transport properties}

\author{Na Hyun Jo}
\affiliation{Ames Laboratory, Iowa State University, Ames, Iowa 50011, USA}
\affiliation{Department of Physics and Astronomy, Iowa State University, Ames, Iowa 50011, USA}

\author{Lin-Lin Wang}
\affiliation{Ames Laboratory, Iowa State University, Ames, Iowa 50011, USA}

\author{Peter P. Orth}
\affiliation{Ames Laboratory, Iowa State University, Ames, Iowa 50011, USA}
\affiliation{Department of Physics and Astronomy, Iowa State University, Ames, Iowa 50011, USA}

\author{Sergey L. Bud'ko}
\affiliation{Ames Laboratory, Iowa State University, Ames, Iowa 50011, USA}
\affiliation{Department of Physics and Astronomy, Iowa State University, Ames, Iowa 50011, USA}

\author{Paul C. Canfield}
\affiliation{Ames Laboratory, Iowa State University, Ames, Iowa 50011, USA}
\affiliation{Department of Physics and Astronomy, Iowa State University, Ames, Iowa 50011, USA}
\email[]{canfield@ameslab.gov}

\date{\today}
\maketitle

{\bf Elastoresistance describes the relative change of a material's resistance when strained. 
It has two major contributions: strain induced geometric and electronic changes.  If the geometric factor dominates, like in ordinary metals such as copper, the elastoresistance is positive and rather small, i.e. typically of order 1.~\cite{gere2001} In a few materials, however, changes in electronic structure dominate, which gives rise to larger and even negative values, such as (-11) for Bi.~\cite{Kuczynski1954} Here, we report that the transition metal dichalcogenide (TMDC) WTe$_{2}$ is a member of the second group, exhibiting a large and non-monotonic elastoresistance that is about (-20) near 100\,K and changes sign at low temperatures. We discover that an applied magnetic field has a dramatic effect on the elastoresistance in WTe$_{2}$: in the quantum regime at low temperatures, it leads to quantum oscillations of the elastoresistance, that ranges between (-80) to 120 within a field range of only half a Tesla. In the semiclassical regime at intermediate temperatures, we find that the elastoresistance rapidly increases and changes sign in a magnetic field. 
We provide a semi-quantitative understanding of our experimental results using a combination of first-principle and analytical low-energy model calculations. Understanding bulk properties of TMDCs under uniaxial strain is an important stepping stone toward strain engineering of 2D TMDCs.} 

The majority of TMDCs have layered structures, which can be easily exfoliated, due to van der Waals bonding between chalcogenides layers.~\cite{Kolobov2016} This low dimensionality, as well as novel electronic transitions and properties, led to studies of the effects of hydrostatic pressure on these materials. As a result, superconductivity was often found upon suppression of charge density wave (CDW) by applying pressure to TMDCs such as TaS$_{2}$, TaSe$_{2}$ and 2H-NbSe$_{2}$.~\cite{manzeli2017} Another particularly striking example is WTe$_{2}$, which displays the fascinating behavior of extremely large magnetoresistance~(XMR)~\cite{Mun2012, Ali2014} at low pressures and superconductivity at higher pressures.~\cite{kang2015} The tunability of TMDCs by hydrostatic pressure, combined with their low dimensionality, suggests that uniaxial strain might be another powerful tuning parameter. Despite the importance of strain for thin films and device applications, a systematic experimental study of the response of bulk TMDCs to continuously controlled strain is still lacking. In addition, first-principle calculations have emphasized that strain can be employed to tune electronic characteristics of TMDCs such as band gap, charge carrier effective masses, thermal conductivity, dielectric properties and spin-orbit coupling.~\cite{manzeli2017} In this work, we demonstrate that WTe$_{2}$ has phenomenal elastoresistance that is highly tunable by an external magnetic field. We employ a combination of quantum oscillation, first-principle density functional theory~\cite{Hohenberg1964,Kohn1965} (DFT) and analytical low-energy model calculations to elucidate the physical origin of this novel "magneto-elastoresistance (MER)" phenomenon. Our analysis provides new insights into the interplay of elastic and electronic degrees of freedom in WTe$_2$ that will facilitate strain engineering of electronic properties of TMDCs in the future. 

WTe$_{2}$ can be considered as a pseudo-1D structure~\cite{Kolobov2016} with distorted zig-zag chains of W atoms running along the $a$ direction. We therefore applied uniaxial strain on bulk single crystals of WTe$_{2}$ along the crystallographic $a$ direction as shown in Fig.\,\ref{fig:1}a and b. We observe that the elastoresistance of WTe$_{2}$, with both strain and electric current directed along the $a$ direction, is very large. It reaches (-20) around a temperature of 100\,K (Fig.\,\ref{fig:1}c). This value is too large to arise from purely geometric changes, considering a Poisson ratio of 0.16.~\cite{Zeng2015} The elastoresistance properties of single crystals of WTe$_{2}$ are thus clearly governed by the strain induced electronic changes. Interestingly, we observe a non-monotonic behavior of elastoresistance that exhibits a rapid increase around 35\,K leading to a sign change at low temperatures. As shown below, this behavior can be understood within a low-energy model calculation as arising from a strain-induced redistribution of carriers among electron and hole pockets with different effective masses. 

WTe$_{2}$ is well-known for its XMR at low temperature. As seen in Fig.\,\ref{fig:2}a, the resistance exhibits a large increase in an applied magnetic field. As shown in Fig.\,\ref{fig:2}b, the temperature dependent elastoresistance also shows a pronounced response to an applied magnetic field. Remarkably, in the quantum regime, at temperatures $T < 12$~K, the elastoresistance changes dramatically with relatively small changes in the magnetic field. The inset of Fig.\,\ref{fig:2} depicts the low-T elastoresistance for $B = 13.1$\,T, 13.6\,T and 14.0\,T. Over this relatively small field range, the elastoresistance varies from 120 to (-75) and back to 50. We show below that such newly observed elastoresistance quantum oscillations occur due to strain tuning of Shubnikov-de Haas (SdH) oscillations (see SI for more detail). Furthermore, at low to intermediate temperatures (e.g. $T\,=\,15$\,K), the elastoresistance changes abruptly as a function of magnetic field from (-10) to (+20) and eventually saturates. Figure\,\ref{fig:2}c  shows a three dimensional plot of the elastoresistance. To avoid the effects of quantum oscillation at the lowest temperatures, we focus on the MER at 15\,K which is highlighted in pink. At this temperature, the elastoresistance changes rapidly below 0.5\,T, and then saturates above 2\,T. This behavior can be readily captured by semiclassical transport calculations, which we will discuss in detail below (and in the SI). Importantly, this discussion shows that measuring MER provides new insights into the strain response of a material (beyond the zero field elastoresistance) as it it probes different strain derivatives. For example, it allows us to infer that the sign change of the elastoresistance in zero field arises from strain tuning of the carrier densities as opposed to tuning of effective masses and scattering rates.

To understand the effect of strain on the electronic properties of WTe$_2$ such as the carrier's effective masses $m^*$ or the cross-sections of extremal orbits on the Fermi surface (FS) (perpendicular to an applied magnetic field), we measure SdH oscillations and use DFT band structure calculations. Figure\,\ref{fig:3}a shows DFT calculations on WTe$_{2}$ without and with strain. Given that the curvature of a band at the Fermi level relates to the effective mass of the charge carrier, one can see that the effective masses change as strain is applied. In addition, the FS areas enclosed by extremal orbits~\cite{Rourke2012} in the $k_{x}$-$k_{y}$ plane at $k_{z}\,=\,0$ also vary as a function of strain as shown in Fig.\,\ref{fig:3}b. 

The changes of the extrema FS area and the effective (cyclotron) masses can be experimentally inferred from analyzing SdH oscillation frequencies and their amplitudes as a function of temperature. After subtracting a quadratic background, the residual oscillatory parts of resistance, with and without stain, are plotted as a function of $1/B$ in Fig.\,\ref{fig:3}c. In order to quantify the FS changes, a Fast Fourier Transform (FFT) was taken in Fig.\,\ref{fig:3}d. All four frequencies shift to higher values when the sample is compressed. Since the frequency is related to the extremal orbit of the FS via the Onsager relation, ${ F }_{ i }=\frac { \hbar c }{ 2\pi e } { S }_{ i }$,\cite{Pippard1989} these positive shifts indicate enlarged extremal FS orbits. Figure\,\ref{fig:3}e shows a linear dependence of frequency on strain. Figure\,\ref{fig:3}c also shows that the amplitude of the SdH oscillation decreases with increasing temperature. Such a tendency allows us to extract the effective masses by fitting with Lifshitz-Kosevich (LK) theory, ${ R }_{ T }=\frac { \alpha { m }^{ * }T/H }{ \sinh(\alpha { m }^{ * }T/H) }$\cite{Pippard1989}, where $\alpha=2\pi^{2} c k_{B}/e \hbar $. Figure\,\ref{fig:3}f demonstrates that all masses increase with compressive strain, even though the magnitude of changes is different, in agreement with our DFT predictions. 

To interpret our MER measurements, we employ an effective low-energy model that can account for the salient experimental features. Using input from DFT calculations as well as ARPES and quantum oscillations measurements, our minimal low-energy model consists of three parabolic bands (see Fig.~\ref{fig:4}): one electron pocket with effective mass $m^*_e$ and two hole pockets with different effective masses $m^*_{lh} \ll m^*_{hh}$. The electron pocket and the light-hole $(lh)$ pocket cross the Fermi energy $E_F$, whereas the heavy-hole ($hh$) pocket sits slightly below $E_F$.~\cite{Pletikosic2014,wu2015,Wu2017}

Let us first discuss the elastoresistivity in zero magnetic field. Within a semiclassical approach, the conductivity is given by $\sigma = \sum_\alpha \sigma_\alpha$ with $\sigma_\alpha = n_\alpha e^2/(\Gamma_\alpha m^*_\alpha)$ being the contributions from individual bands. Here, $n_\alpha$ is the carrier density and $\Gamma_\alpha$ is the scattering rate of band $\alpha$. The zero-field elastoresistivity thus reads
\begin{equation}
\label{eq:1}
\frac{1}{\rho(0)} \frac{d \rho(0)}{d \epsilon} = \sum_\alpha \frac{\sigma_{\alpha}(0)}{\sigma(0)} \left[\frac{\zeta^{(\alpha)}_m}{m^{*}_\alpha} + \frac{\zeta^{(\alpha)}_\Gamma}{\Gamma_\alpha} - \frac{\zeta^{(\alpha)}_n}{n_\alpha}\right] \,,
\end{equation}
where $\rho(0) \equiv \rho(\mathbf{B} = 0)$ and we have introduced the strain derivatives $\zeta^{(\alpha)}_m = \frac{d m^*_\alpha}{d \epsilon}$, $\zeta^{(\alpha)}_\Gamma = \frac{d \Gamma_\alpha}{d \epsilon}$ and $\zeta^{(\alpha)}_n = \frac{d n_\alpha}{d \epsilon}$. Increasing $m^*_\alpha$ and $\Gamma_\alpha$ increases $\rho$, whereas increasing the carrier density $n_\alpha$ reduces $\rho$. Contributions from different bands are weighted according to their contribution to the total conductivity. One can estimate the strain response of the scattering rates $d \Gamma_\alpha/d\varepsilon$ using expressions for impurity and phonon scattering rates from Boltzmann theory and using Matthiessen's rule~\cite{SmithJensen-Transport-Book} (see also the SI). 

At low temperatures, where impurity scattering dominates, the elastoresistivity then becomes $\frac{1}{\rho(0)} \frac{d\rho(0)}{d \epsilon} = 2 \sum_{\alpha} \frac{\sigma_\alpha}{\sigma} \bigl( \frac{\zeta^{(\alpha)}_m}{m^*_\alpha} - \frac{1}{3}\frac{\zeta^{(\alpha)}_{n}}{n_\alpha}\bigr)$. From analyzing quantum oscillations (see Fig.~\ref{fig:3}), we conclude that both strain derivatives have the same sign $\zeta^{(\alpha)}_m, \zeta^{(\alpha)}_{n} < 0$ such that the two effects compete with each other. Which of the two dominates depends on microscopic details. A DFT transport calculation, which keeps the scattering rates constant, predicts a positive elastoresistivity, in agreement with experimental results (see SI for details). As shown in Fig.~\ref{fig:4}, this behavior is also readily captured within our effective three-band model, where we use as input from DFT that the two holes bands move up in energy under compressive strain, while the electron band moves down (see Fig.~\ref{fig:3}a). Interestingly, the heavy-hole band shifts up by an amount about ten times larger than the other two bands, yet still remains below $E_F$. At low temperatures, it thus remains completely filled and does not contribute to transport. Electrons redistribute solely among the electron and light-hole pockets such that both $n_e$ and $n_{lh}$ increase, causing an increase in the total carrier density $n$ and positive elastoresistivity (see Fig.~\ref{fig:4}d). In the experiment, this effect seems to dominate over the strain-induced enhancement of the effective masses, which tends to reduce the elastoresistivity. 

The situation is different at finite temperature, where one needs to take into account additional bands that lie within a range of $k_B T$ of the Fermi energy. The non-monotonic behavior and associated sign change of the elastoresistance as a function of $T$ can thus be understood as a result of a redistribution of carriers including those nearby bands, i.e. including the $hh$ band. A precise modeling of the strain-induced modifications of the bandstructure at finite $T$ require further experimental studies, e.g. ARPES under finite strain, or first-principle calculations that take thermal expansion effects into account. We can, however, capture the experimentally observed behavior of Fig.~\ref{fig:1} within our low-energy model. As shown in Fig.~\ref{fig:4}, the decrease of the elastoresistivity (as $T$ increases) follows from the fact that the heavy-hole pocket is shifted upwards by strain to within a range of $k_B T$ of the Fermi energy. It is thus only partially occupied at finite $T$, and holes redistribute from the light-hole to the heavy-hole pocket. This leads to the observed increase of the resistivity at finite temperatures, and the eventual sign change. Note that there is an additional negative contribution to elastoresistivity from the enhancement of the effective masses. The broad minimum of the elastoresistivity arises within our model from a competition of electrons moving from the heavy-hole pocket to both the light-hole band (decreasing $n_{lh}$ and $\rho$) and the electron band (increasing $n_e$ and $\rho$). Specifically, the slope of the elastoresistivity is determined by the ratio of the density of states $g_{lh}/g_e$ at the chemical potential $\mu(T)$. The slope is negative for $g_{lh} > g_{e}$, as electrons from the heavy-hole pocket are then more likely to enter the light-hole one. The precise value of this ratio depends on microscopic details. 

At finite magnetic field, the MER shows distinct behavior in the low-temperature quantum regime $T \lesssim 15$~K and the semiclassical regime at higher $T$. In the quantum regime, the sensitivity of the MER to strain and fields described above, readily follows from strain tuning of the Shubnikov-de Haas (SdH) oscillation frequencies. 
In the semiclassical regime $T \gtrsim 15~\text{K}$, the observed MER can be captured by an effective two-band model of electron and hole carriers with density $n_e, n_h$ and mobilities $\mu_e=e/(m^*_e \Gamma_e)$ and $\mu_h$. We capture the contribution of the two hole pockets via an effective hole mobility $\mu_h=e/(\bar{m}^*_h \bar{\Gamma}_h)$, where $\bar{m}^*_h, \bar{\Gamma}_h$ are averages of the two hole pockets. As shown in detail in the SI, we start from the well-known semiclassical form of the resistivity $\rho(B) = \rho(0) \left( \frac{1 + e \rho(0) (n_e \mu_h + n_h \mu_e) \mu_e \mu_h B^2}{1 + [e \rho(0) \mu_e \mu_h (n_e - n_h)  B]^2}\right)$ and consider ``intermediate'' magnetic field strengths, where the magnetoresistance is not yet saturated, to arrive at 
\begin{equation}
\label{eq:2}
\frac{1}{\rho(B)} \frac{d \rho(B)}{d \epsilon} - \frac{1}{\rho(0)} \frac{d \rho(0)}{d \epsilon} =  2 \mu^2 B^2 \frac{\rho(0)}{\rho(B)} \sum_{\alpha} \frac{\sigma_\alpha}{\sigma(0)} \left( - \frac{\zeta^{(\bar{\alpha})}_m}{m^*_{\bar{\alpha}}} - \frac{\zeta^{(\bar{\alpha})}_{\Gamma}}{\Gamma_{\bar{\alpha}}}\right) \,.
\end{equation}
Here, $\bar{\alpha}= e (h)$ for $\alpha = h (e)$, and we have approximated $\mu_e \approx \mu_h \equiv \mu$ (see the SI for the  general case). Importantly, we find a deviation of the finite field elastoresistivity from the zero-field one that increases as $B^2$ for small fields $B \ll B_1 = e \rho(0) [(n_e \mu_h + n_h \mu_e) \mu_e \mu_h]^{-1/2}$. It reaches saturation when $B \gg B_1$ due to the term $\rho(B)$ in the denominator (see Fig.~\ref{fig:2}(c) and SI). Note that $B_1$ can also be extracted from the coefficient of the non-saturating, quadratic MR $= (B/B_1)^2$, which yields that $B_1$ increases from $B_1(30~\text{K}) = 1.4$~T to $B_1(90~\text{K}) = 12$~T,~\cite{Ali2014} in agreement with our findings. According the Eq.~\eqref{eq:2} the prefactor of the quadratic MER term depends on the strain response of the effective masses and the scattering rate, but interestingly does not depend on the change of the carrier density. It probes a different strain derivative and thus provides novel insights beyond a zero-field elastoresistivity measurement. In particular, unlike, the elastoresistivity in zero field, it exhibits the same sign for all temperatures. This allows us to conclude that the sign change of the $B=0$ elastoresistivity is rooted in the non-monotonic behavior of the strain derivative of the carrier densities $\zeta^{(\alpha)}_n(T)$ (see Fig.~\ref{fig:4}). Finally, from our effective model we find that $\zeta^{(\alpha)}_m < 0$ and $\zeta^{(\alpha)}_\Gamma < 0$, and therefore predict a quadratically increasing MER (at all $T$ and low $B \ll B_1$), in agreement with experiment (see Fig.~\ref{fig:2} and SI Fig.~9).

To conclude, we report a large and non-monotonic MER in high-mobility crystals of WTe$_2$. This provides new insights into the strong coupling of elastic and electronic degrees of freedom in this material, opening a route to engineer magneto-transport properties via strain. Using a combination of density-functional theory and analytical low-energy multi-band model calculations, we relate these observations to strain-induced changes of the bandstructure, resulting in an increase as well as a redistribution of carriers between bands with different mobilities. Promising future research directions are to investigate the impact of thermal expansion and electronic correlations on both the bandstructure and its response to strain, as well as the role of the edge states in magnetotransport under finite strain. 

\clearpage

\section{Methods}

\begin{figure}
	\includegraphics[scale=1]{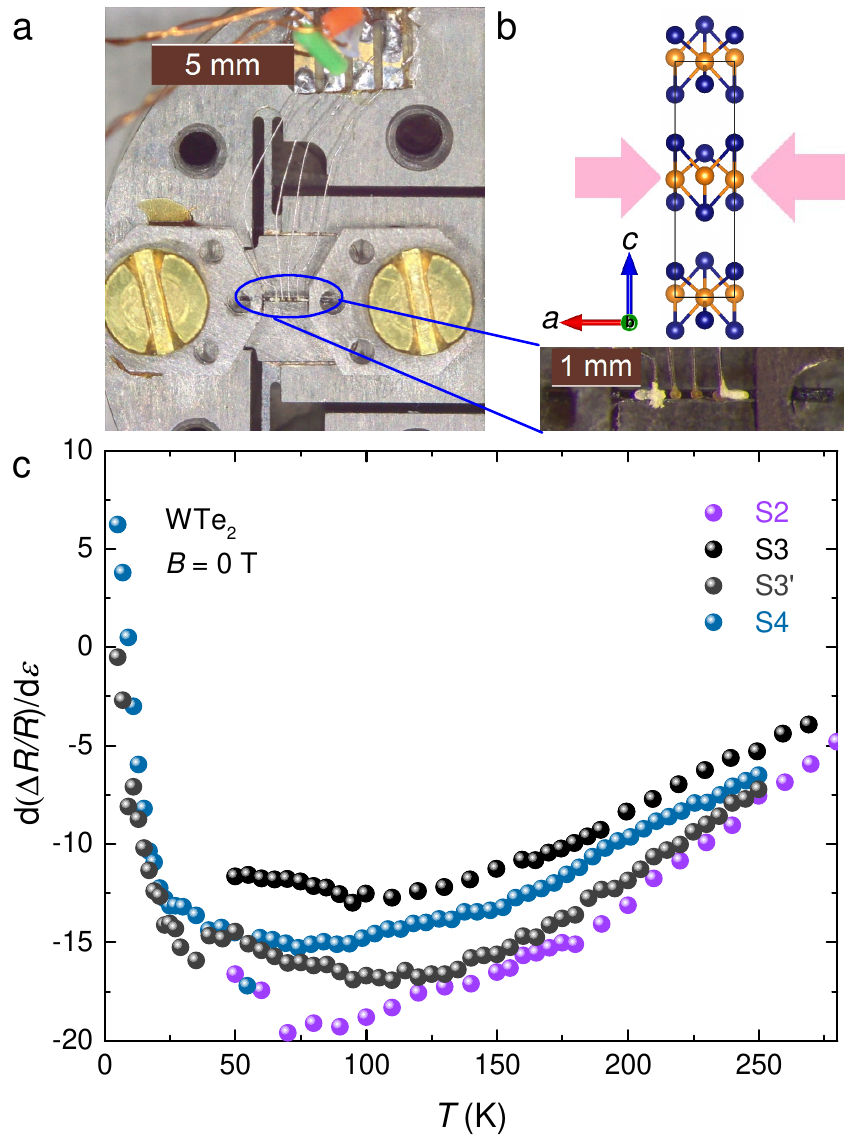}%
	\caption{\textbf{Measurement setup and zero field elastoresistance.} \textbf{a}, A single crystal of WTe$_{2}$ is mounted on Razorbill CS100 cryogenic uniaxial strain cell. The electrical current and mechanical stress was applied along the crystallographic $a$ direction, and the magnetic field was applied along the crystallographic $c$ direction. \textbf{b}, Crystal structure of WTe$_{2}$. \textbf{c}, Elastoresistance of WTe$_{2}$ in the temperature range of $5\,\textrm{K}\,\le\,T\,\le\,270\,\textrm{K}$ with $B\,=\,0$\,T for samples S2, S3, S3' and S4. All samples exhibit a large, negative elastoresistance at not too low temperatures. S3 and S4 show a pronounced upturn at low temperatures $T \lesssim 25$~K. Quantitative differences between samples can be due to a difference in quality and/or strain transmission through the sample.}
	\label{fig:1}
\end{figure}

\begin{figure}
	\includegraphics[width=6.5in]{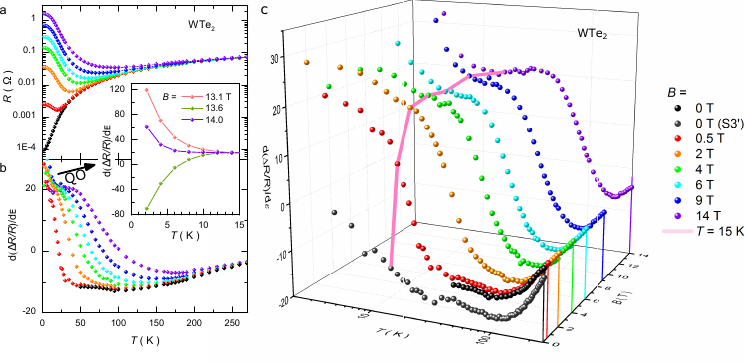}%
	\caption{\textbf{Magnetoresistance (MR) and magneto-elastoresistance (MER)} \textbf{a}, Resistance of sample S3 as a function of temperature for various magnetic fields, $H = 0 , 0.5 , 2, 4 , 6 , 9, 14$ (in units of T). The legend is shown on the right. The field is applied along the crystallographic $c$ direction. \textbf{b}, Elastoresistance of sample S3 as a function of temperature in the same magnetic fields $H$. Inset shows elastoresistance in the low temperature regime ($2\,\textrm{K}\,\le\,T\,\le\,15\,\textrm{K}$) with an applied magnetic field of 13.1\,T, 13.6\,T and 14\,T.\textbf{c}, Three-dimensional representation of MER data from panel (b) including $H= 0$T data of sample S3'. The pink line connects the data at $T\,=\,15$\,K, highlighting the rapid increase and eventual saturation of MER in  magnetic field. }
	\label{fig:2}
\end{figure}

\begin{figure*}[tb]
	\includegraphics[width=6.5in]{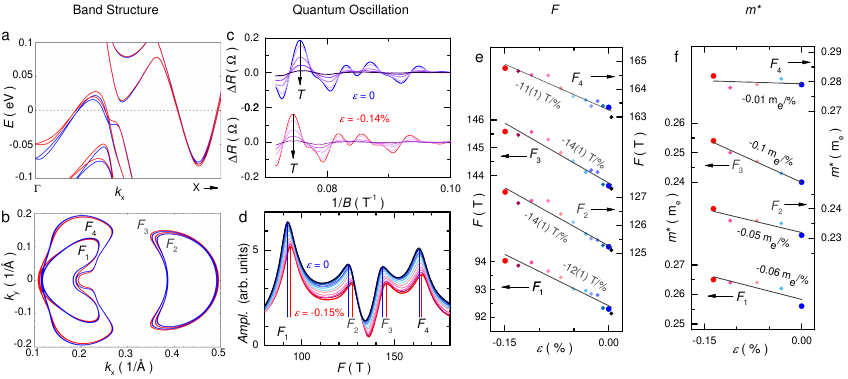}%
	\caption{\textbf{DFT results and quantum oscillation analysis under uniaxial strain.} \textbf{a}, Results of DFT band structure calculation along the $\Gamma$\,-\,$X$ direction without strain ($\epsilon = 0\%$, blue) and with strain ($\epsilon=-0.2\%$, red). \textbf{b}, Strain induced modification of extremal orbits at $k_{z}\,=\,0$ from DFT calculation. Blue and red lines refer to the same strain as in panel (a). Fermi surfaces $F_{1}$ and $F_{4}$ correspond to hole bands and $F_{2}$ and $F_{3}$ to electron bands. \textbf{c}, Experimental results for Shubnikov-de Haas (SdH) quantum oscillations after subtracting the background at temperatures $T= 2, 3, 5, 7 and 10$~K. Upper panel is for $\epsilon\,=\,$0\,$\%$ and lower panel is for $\epsilon\,=\,$-0.14\,$\%$. \textbf{d}, Fast Fourier transformation (FFT) results at various strains. Blue lines indicate the frequencies at zero strain and red lines are located at the frequencies at -0.15\,$\%$ strain. \textbf{e}, SdH oscillation frequencies of the four extremal orbits $F_{1,2,3,4}$ as a function of strain $\epsilon$. The numerical values of the slopes are given in the figure. \textbf{f}, Effective cyclotron masses of the four extremal orbits $F_{1,2,3,4}$ as a function of strain with slopes given in the figure.}
	\label{fig:3}
\end{figure*}

\begin{figure*}[h!]
	\includegraphics[width=.9\textwidth]{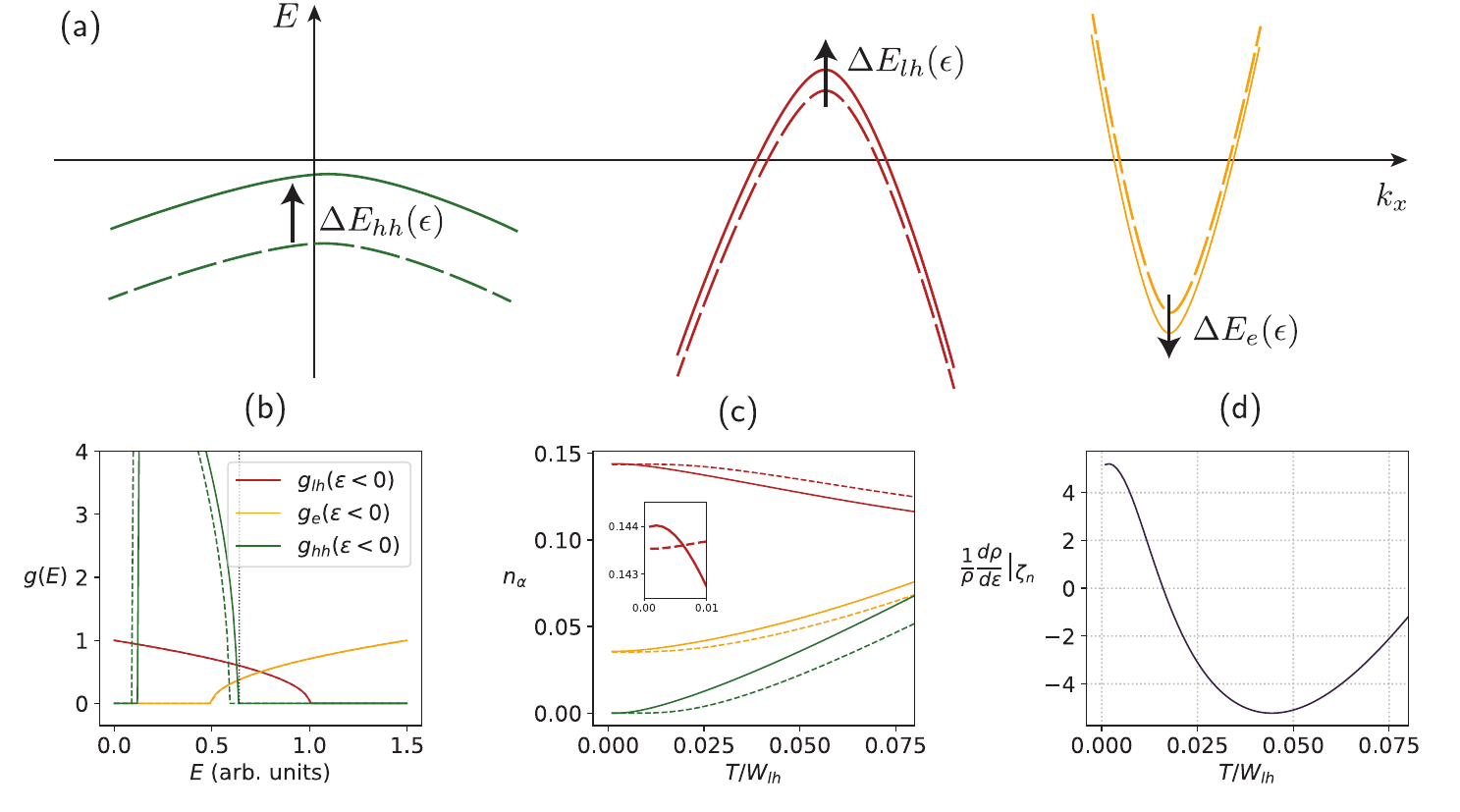}%
	\caption{
		\textbf{Theoretical low-energy model analysis} (a) Schematics of electron and hole pockets in low-energy three-band model before (dashed) and after (solid) application of compressive strain. Direction and relative size of strain-induced band shifts $\Delta E(\epsilon)$ are taken from DFT calculation (see Fig.~\ref{fig:3}).
		(b) Density of states $g(E)$ of three-band model before (dashed) and after (solid) rigid band shifts $\Delta E_{e}= - 2 \times 10^{-3}$ and $\Delta E_{\text{hh}} = 4 \times 10^{-2}$. All energies are in units of the light-hole bandwidth $W_{lh}$. Bandwidths are set to $W_e = W_{lh} = 1$ and $W_{hh} = 0.5$. Effective masses assumed to be strain-independent for simplicity, and their ratios set to $m^*_{e}/m^*_{lh}=1$ and $m^*_{hh}/m^*_{lh} = 4.6$. The bottom of the bands are at $E^{\text{min}}_{lh}=0$, $E^{\text{min}}_{e} = 0.5$, and $E^{\text{min}}_{hh} = 0.09$. The chemical potential $\mu(T=0) = 0.64 \, W_{lh}$ is indicated by the vertical dotted line. 
		(c) Carrier densities of the different bands $n_\alpha$ as a function of temperature $T$ for zero and finite strain. Colors and linestyles are identical to panels (a) and (b). Inset shows that $n_{lh}$ increases at low $T$ (as long as $E_{hh}(\varepsilon) - \mu \ll T$), but decreases at higher $T$, causing the sign change of the ER. 
		(d) Elastoresistivity that arises from redistribution of carriers due to rigid band shifts $\Delta E(\epsilon)$ shown in panel (b). We have assumed that these band shifts are caused by strain of size $\epsilon=10^{-3}$. This is justified from DFT calculations, which predicts bandshifts of $2-30$~meV (for the different bands) for $\epsilon=10^{-3}$. Note that the $T$-axis scale is of the same order, corresponding to a range from zero to roughly (two times) room temperature ($\Delta E_{hh} = 0.04 \approx 30$meV). At $T=0$, $ER> 0$ as electrons move from the lh to the e band, increasing the total number of carriers (see inset in panel (c)). At $T>0$, the hh band is only partially filled and moving it closer to the chemical potential shifts hole carriers from $lh$ to $hh$ pocket. We note that the total ER also contains a contribution from the strain-induced increase of the effective masses, which leads to an approximately $T$-independent negative shift of ER.}		
	\label{fig:4}
\end{figure*}

Single crystals of WTe$_2$ were grown by solution method, following the procedure described in Ref.\,~\onlinecite{wu2015}. Temperature and field dependent transport properties were measured in a Quantum Design(QD), Physical Property Measurement System for 1.8\,$\le T \le$\,300\,K and $\left| H \right| \le$ \,140\,kOe. Resistance measurements under uniaxial strain were carried out using a Razorbill CS100 cryogenic uniaxial strain cell. To be more specific, outer and inner piezoelectric stacks were controlled by two Keithley Model 2400 source meters, and corresponding length changes of the crystals were measured by a capacitance sensor using an Andeen-Hagerling(AH) Model 2550A capacitance bridge. Note that there is always certain errors in the length change measurements due to thermal contraction of various parts. Although some thermal effects were addressed (i.e. the thermal effect from piezoelectric materials was canceled via symmetric usage of outer and inner piezoelectric stacks and the thermal effect from the capacitance sensor was  removed by subtracting the temperature dependence capacitance of the empty cell result), others, including the thermal contraction from the cell and the crystal itself, were not considered. The crystals were mounted across the two plates with Stycast 2850 FT, so that the crystal was mechanically attached to the plates firmly and electrically isolated from the cell body.(Fig.\,\ref{fig:1}a) We estimated that, due to the epoxy, about 80\,$\%$ of the displacement was transmitted as sample strain. In this calculation, we used the Young's modulus of the crystals of 80\,GPa and the thickness of the glue was $\sim\,50\,\mu m$.~\cite{Brodsky2017,Lee2016} The contacts for the electrical transport measurement were prepared in a standard linear four-probe configuration using Epotek-H20E silver epoxy and silver paint, and Lakeshore Model 370 AC resistance bridge conducted the resistance measurement of the crystals. Due to the large drop in the resistivity of WTe$_{2}$ upon cooling in zero field combined with the limit of the resistance bridge (resolution is 1\,$\mu\ohm$ with $I\,=\,3.16$\,$\textrm{mA}$ in the range of 2.0\,m$\ohm$), we were not able to conduct the elastoresistance measurements below $T\,<\,50\,\textrm{K}$ with $H\,=\,0\,\textrm{T}$ in the above experimental setting. In order to overcome this limit of the measurement, we performed the elastoresistance experiment without the magnetic field using Stanford Research Systems(SRS) 860, Lock-In Amplifier and SRS Model CS580, voltage controlled current source in a Janis SHI-950-T closed cycle cryostat. Two samples were measured in this way, S3' and S4. S3' is esentially same as S3, which was measured in PPMS, but new contacts were made after cleaving the top layer. S4 was mounted in Razorbill CS130 cryogenic uniaxial strain cell. S4 was secured with a small amount of Devcon 5 minute epoxy in between anodized plates, which can give the largest transmitted strain on the sample. 

Strain is defined as $\epsilon\,[\%]\,=\,(L-L_{0})/L_{0}\,\times\,100$, where $L_{0}$ is the unstrained length. Thus, a positive sign represents tensile strain and a negative sign stands for compressive strain. We noticed that the crystals are very easy to break when even a small amount of tensile strain is applied. The compressive strain at which the sample starts to buckle can be calculated based on the ratio between length and thickness ($L/t$) of the crystals: $L/t\,=\,\pi/\sqrt{3\epsilon}$.~\cite{Hicks2014} From the calculation, we expected buckling of all three samples that we measured above $\epsilon\,=\,-\,1.5\,\%$. However, the first sample of WTe$_{2}$, S1, cleaved at $\epsilon\,\sim\,-\,0.3\,\%$, which was  before the sample started to buckle due to the easily exfoliatable nature of the crystal. In addition, the second crystal, S2, showed a jump at $\epsilon\,\sim\,-\,0.16\,\%$ without visual observation of a cleave or crack in the sample. This might indicate that cracks or small cleaving can happen even with small strain due to a layered structure. Based on all of the above, elastoresistance measurements were done only $\epsilon\,\le\,\pm 0.013$\,$\%$ which corresponds to a maximum voltage applied to the piezoelectric material of $V\,=\,\pm 5$\,V for the third single crystals of WTe$_{2}$, S3. Within this range, resistance response to the strain ($\Delta R/R$ Vs. strain) was linear without hysteresis (more details are in SI). Details about Shubnikov de Haas oscillations can be found in SI.    

Band structures of WTe$_{2}$ at strains from -0.2 to 0$\%$ were calculated in density functional theory~\cite{Hohenberg1964,Kohn1965} (DFT) using local density approximation~\cite{Ceperley1980,Perdew1981} (LDA) with spin-orbit coupling (SOC) effect included. The dimensions of the unit cells were determined from experimental lattice constants~\cite{Mar1992} ($a$\,=\,3.477\,$\AA$, $b$\,\,=6.249\,$\AA$ and $c$\,=\,14.018\,$\AA$) plus strain and a Poisson ratio~\cite{Zeng2015} of 0.16. The ionic positions in the unit cells were relaxed at different strains and then band structures are calculated. The carrier densities were calculated from the volume of electron and hole pockets in reciprocal space. The quantum oscillation frequencies were calculated by finding the extreme orbit~\cite{Rourke2012} of hole and electron pockets with the magnetic field along the $c$ direction. The conductivity without magnetic field were calculated from the semi-classical Boltzmann equation with the interpolated DFT band structures.~\cite{Madsen2006} DFT calculations were done in VASP~\cite{Kresse1996} with a plane-wave basis set and projector augmented wave~\cite{Bloechl1994} method. We used the orthorhombic cell of 12 atoms with a $\Gamma$-centered Monkhorst-Pack~\cite{Monkhorst1976} (12\,$\times$\,6\,$\times$\,3) $k$-point mesh. The kinetic energy cutoff was 223\,eV. The convergence with respect to $k$-point mesh was carefully checked, with total energy converged, e.g., well below 1\,meV/atom. For ionic relaxation, the absolute magnitude of force on each atom was reduced below 0.01\,eV/$\AA$. 

\begin{acknowledgements}
The authors acknowledge G. Drachuck for technical support, in particular with the computer program used for elastoresistance data collection. The authors thank R. McDonald, E. J. Koenig, A. Kaminski, J. S. Van Dyke, A. Kreyssig, E. Gati, W. R. Meier and Y. Wu for helpful discussion. Research was supported by the U.S. Department of Energy, Office of Basic Energy Sciences, Division of Materials Sciences and Engineering. Ames Laboratory is operated for the U.S. Department of Energy by the Iowa State University under Contract No. DE-AC02-07CH11358. Na Hyun Jo is supported by the Gordon and Betty Moore Foundation EPiQS Initiative (Grant No. GBMF4411). P.P.O. acknowledges support from Iowa State University Startup Funds. 
\end{acknowledgements}

\clearpage

\clearpage

\section{Supplementary Information}

\subsection{\texorpdfstring{Elastoresistance tensor and Gauge factor}{space}}

The elastoresistance of a material characterizes the changes of the resistance of the material due to the stress experienced by the material. Based on the definition of resistance, $R\,=\,\rho\frac{L}{A}$, changes in the resistance can be written as 
\begin{equation}
\frac{dR}{R}\,=\,\frac{d\rho}{\rho}+\frac{dL}{L}-\frac{dA}{A}.
\end{equation}
The first term is change in the resistivity of the material which relates to the electronic properties of the material. The second and the third terms are purely geometrical factors. For typical metals, these geometric terms dominate elastoresistance. However, strain dependent changes in density of states, mobility, scattering, etc. (elastoresistivity) play important roles in cases like WTe$_{2}$.

The elastoresistivity is described by 
\begin{equation}
(\frac { d\rho  }{ \rho  } )_{ i }\, =\,\sum _{ k=1 }^{ 6 }{ { m }_{ ik }\epsilon _{ k } }
\end{equation}
where $m_{ik}$ is the elastoresistive strain matrix and $\epsilon_{k}$ is the strain tensor in Voigt notation.~\cite{Kuo2013} For an orthorhombic structure, there are nine independent terms in the elastoresistivity tensor:~\cite{newnham2005}
\begin{equation}
\begin{pmatrix} \begin{matrix} m_{11} & m_{12} & m_{13} \\ m_{12} & m_{22} & m_{23} \\ m_{13} & m_{23} & m_{33} \end{matrix} & \begin{matrix} 0 & 0 & 0 \\ 0 & 0 & 0 \\ 0 & 0 & 0 \end{matrix} \\ \begin{matrix} 0 & 0 & 0 \\ 0 & 0 & 0 \\ 0 & 0 & 0 \end{matrix} & \begin{matrix} m_{44} & 0 & 0 \\ 0 & m_{55} & 0 \\ 0 & 0 & m_{66} \end{matrix} \end{pmatrix}.
\end{equation}

Considering our measurement configuration, the stress was applied only along the crystallographic $a$ direction, the stress is
\begin{equation}
\tau\,=\,(\tau_{xx},\tau_{yy},\tau_{zz},\tau_{yz},\tau_{zx},\tau_{xy})\,=(\tau_{xx},0,0,0,0,0).
\end{equation}
Since the stain $\epsilon$ can be expressed in terms of the elastic compliance $S$ and the stress $\tau$, 
\begin{equation}
\epsilon_{k}\,=\,\sum _{ l=1 }^{ 6 }{ {S}_{ kl }{ \tau  }_{ l } }, 
\end{equation}
the strain therefore is: 
\begin{equation}
\epsilon\,=\,(\epsilon_{xx},\epsilon_{yy},\epsilon_{zz},0,0,0).
\end{equation}
If we neglect geometric factors, the change in resistance of the crystal is given by 
\begin{equation}
(\frac { dR  }{ R  } )_{ xx }\, =\,m_{11}\epsilon_{xx}+m_{12}\epsilon_{yy}+m_{13}\epsilon_{zz}
\end{equation}
The strain terms in the $y$ and $z$ directions are determined by the Poisson's ratio of the crystal, $\nu_{xy}$ and $\nu_{xz}$.
\begin{equation}
(\frac { dR  }{ R  } )_{ xx }\, =\,\epsilon_{xx}(m_{11}+\nu_{xy}m_{12}+\nu_{xz}m_{13}).
\end{equation}

The Gauge factor (GF) can also be obtained from Eq. (1): 
\begin{equation}
GF\,=\,(\frac { (dR/R)_{xx}  }{ \epsilon_{xx}  } )\,=\,\frac { (d\rho/\rho)_{xx}  }{ \epsilon_{xx}  } + (1+2\nu) .
\end{equation}
The GF of ordinary metals, like copper, is a temperature independent value with magnitude around 2. This is because geometric factor is the dominant term and Poisson's ratio for the most of metals is $0.3\,<\,\nu\,<0.5$.~\cite{gere2001} A recent paper on WTe$_{2}$ calculated the Poisson's ratio $\nu$ of the material as $\nu\,\sim\,0.16$.~\cite{Zeng2015,Lee2016} Since we obtained GFs much larger than 2 for WTe$_2$, changes in the electronic properties rather than geometry, are dominant. Therefore, measurements of elastoresistivity (and its response to magnetic field) offer new insight into the electronic properties of WTe$_{2}$. 

\subsection{\texorpdfstring{Elastoresistance experiment}{space}}

\begin{figure}
	\includegraphics[scale=1]{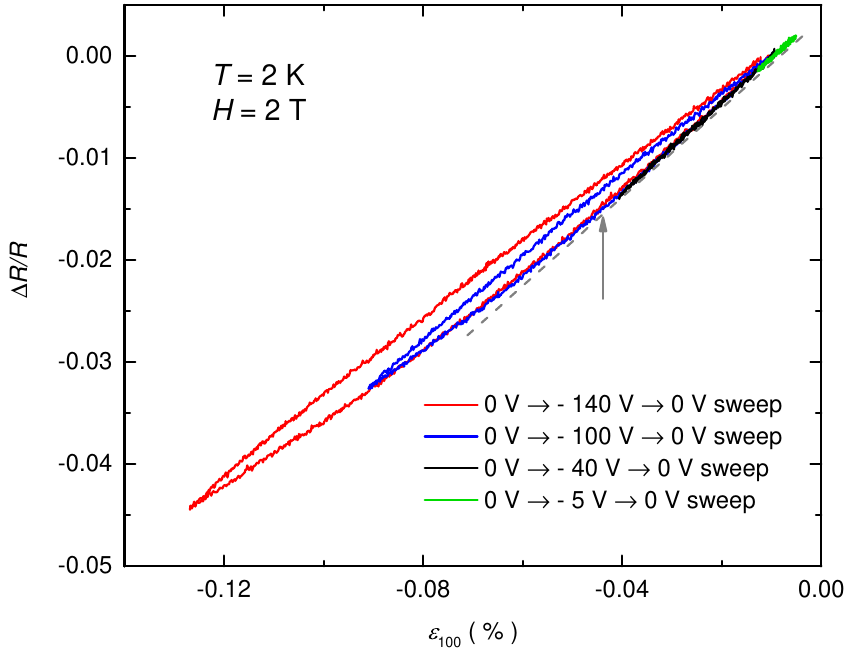}%
	\caption{\textbf{Resistance response to the strain with various voltage limits} Strain($\epsilon_{100}$) sweep at fixed temperature of $T$\,=\,2\,K and the magnetic field $H$\,=\,2\,T with various voltage limits; -5 V (green line), -40 V (black line), -100 V (blue line) and -140 V (red line). The Gray dashed line is a linear guide line to point out the deviation from linear behavior, which also indicated with the gray arrow.    
		\label{Hsweep}}
\end{figure}

\begin{figure}
	\includegraphics[scale=1]{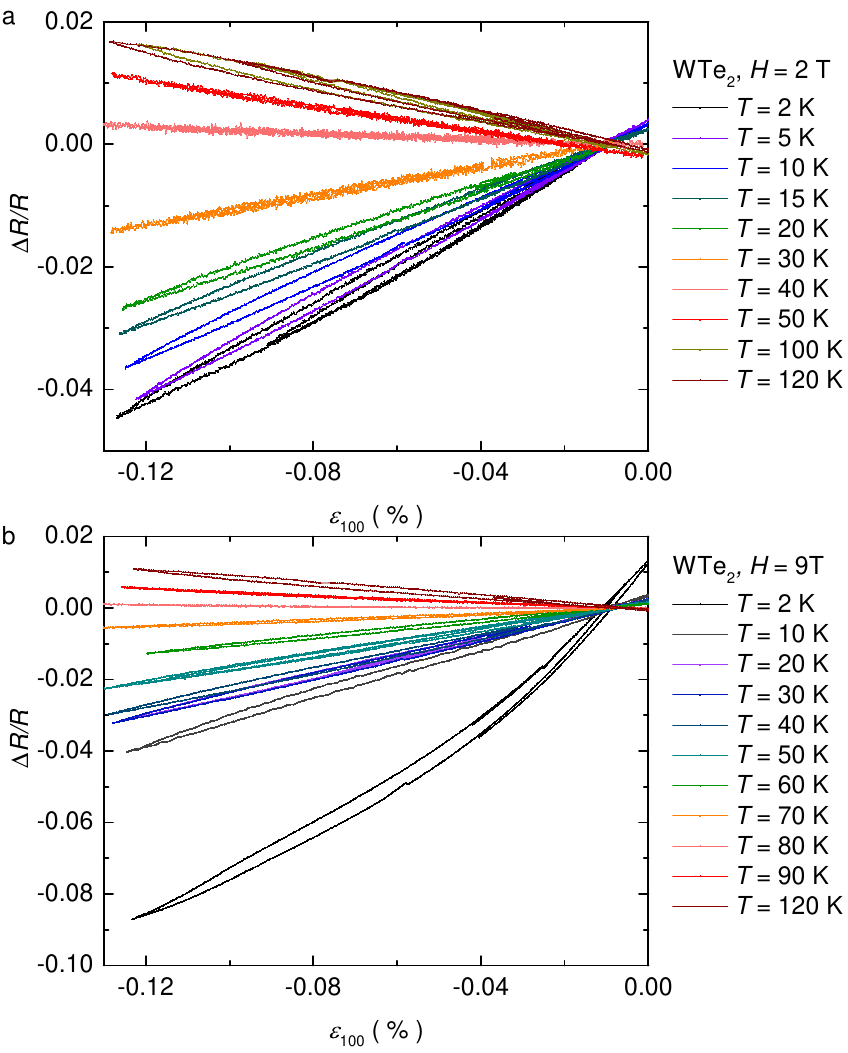}%
	\caption{\textbf{Resistance response to the strain with various temperatures} \textbf{a}, Strain($\epsilon_{100}$) sweep at the magnetic field of 2\,T. \textbf{b}, Strain($\epsilon_{100}$) sweep at the magnetic field of 9\,T. Note: clear non-linearity in 2\,K data is likely associated with strain induced changes in QO frequencies and QO in MER is discussed below.
		\label{HsweepL}}
\end{figure}

We took care to measure our data in the linear response limit.  To do this we measured $\Delta R/R$ versus strain loops for a variety of strain (or voltage) sweeps at base temperature (Fig.\,\ref{Hsweep}) as well as over the whole temperature range (Fig\,\ref{HsweepL}).  We find that as we increase the size of strain sweeps there is an increasing hysterisis.  As a result of these measurements and out of an abundance of caution, we limit our measurements to $\left| \epsilon \right| \,\le\,0.013\,\%$ so as to be in the linear regime.  

\subsection{\texorpdfstring{Large changes in elastoresistance due to quantum oscillations}{space}}

\begin{figure}
	\includegraphics[scale=1]{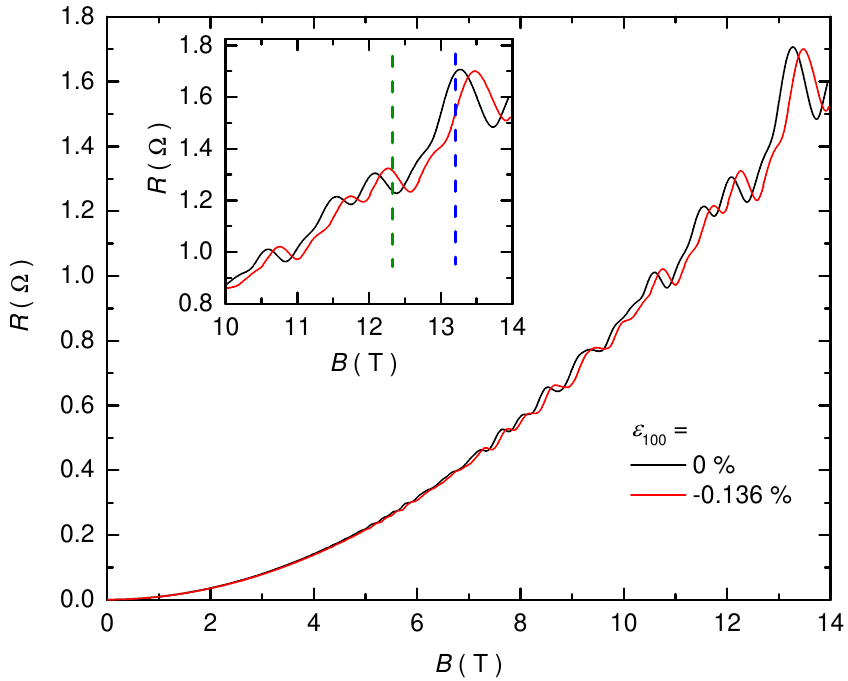}%
	\caption{\textbf{Large changes in elastoresistance} Magnetoresistance data without(black line, $\epsilon\,=\,0\,\%$) strain and with (red line, $\epsilon\,=\,-0.136\,\%$) at $T\,=\,2$\,K. Inset shows enlarged magnetoresistances from 10\,T to 14\,T.      
		\label{QOeffect}}
\end{figure}

Figure\,\ref{QOeffect} shows magnetoresistance without and with ($\epsilon\,=\,-0.136\,\%$) strain at $T\,=\,2$\,K. Whereas the MR increases quadratically without saturation up to 14\,T, quantum oscillations are detected above $\sim$\,4\,T in both cases. Due to the strain induced changes in frequencies, there are mismatches in quantum oscillation peaks in the high magnetic field regime. Therefore, elastoresistances change dramatically from negative to positive with small changes of the magnetic field. To illustrate this more clearly, enlarged magnetoresistances from 10\,T to 14\,T are shown in the inset of Fig.\,\ref{QOeffect}. The green dashed line indicates negative elastoresistance whereas the blue dashed line indicates positive elastoresistance. 

\begin{figure}
	\includegraphics[scale=1.7]{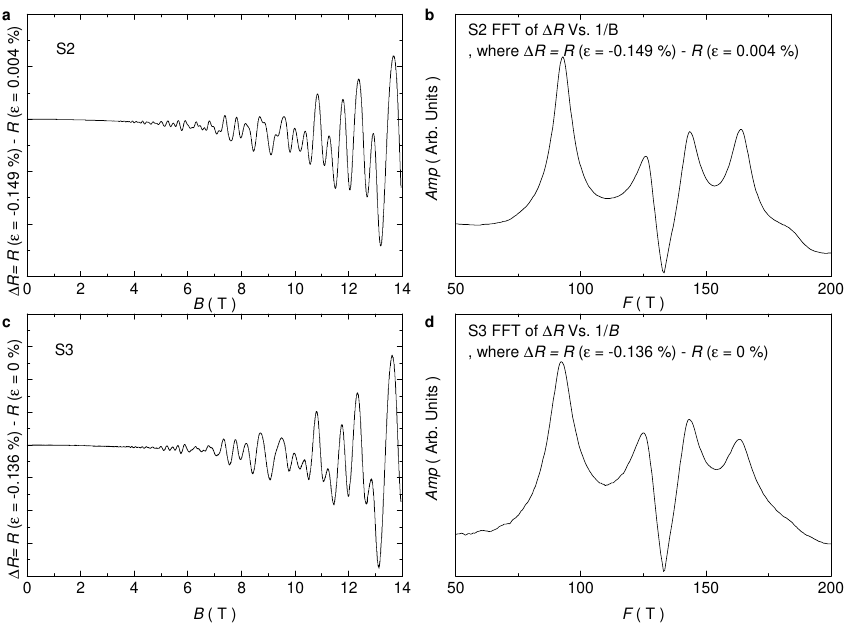}%
	\caption{\textbf{QO in MER} \textbf{a}, $\Delta\,R\,=\,R\,(\epsilon\,=\,-0.149\,\%)\,-\,R\,(\epsilon\,=\,0.004\,\%)$ as a function of the magnetic field in S2.\textbf{b}, FFT of $\Delta\,R$ in terms of 1/$B$ in S2. \textbf{c}, $\Delta\,R\,=\,R\,(\epsilon\,=\,-0.136\,\%)\,-\,R\,(\epsilon\,=\,0\,\%)$ as a function of the magnetic field in S3. \textbf{d}, FFT of $\Delta\,R$ in terms of 1/$B$ in S3.   
		\label{QOinER}}
\end{figure}

Subtracting the MR without strain from MR with strain allows us to directly examine oscillations in the MER as shown in Fig.\,\ref{QOinER}a and c. In order to confirm these as quantum oscillations, we did FFT on the data with 1/$B$. Figure\,\ref{QOinER}b and d are the results of FFT on S2 and S3. All four frequencies, that were detected from SdH, are observed in both samples. As such, these data, as well as the data highlighted in the inset of Fig.\,2a and b, are clear manifestations of quantum oscillations in MER. 

Based on Lifshitz and Kosevich fomula, one can write MER oscillation as below.
\begin{equation}
MER_{osc}\,=\,\sum _{ i }{ { A }_{ i }\left[ sin\left\{ \frac { 2\pi ({ F }_{ i }+\delta { F }_{ i }) }{ B } +{ \phi  }_{ i } \right\} -sin\left\{ \frac { 2\pi { F }_{ i } }{ B } { +\phi  }_{ i } \right\}  \right] /\varepsilon  } .
\end{equation}

For $\delta F\,\ll \,F$ and $\frac { 2\pi \delta { F }_{ i } }{ B } \ll 1$,
\begin{equation}
MER_{osc}\,=\,\sum _{ i }{ { A }_{ i }cos\left\{ \frac { 2\pi { F }_{ i } }{ B } +{ \phi  }_{ i } \right\} \cdot \frac { 2\pi { \delta F }_{ i } }{ B } \cdot \frac { 1 }{ \varepsilon  }  } \quad \quad 
\end{equation}

Note that the frequency of oscillations are same as SdH. 

\subsection{\texorpdfstring{Quantum oscillation analysis}{space}}

\begin{figure}
	\includegraphics[scale=1]{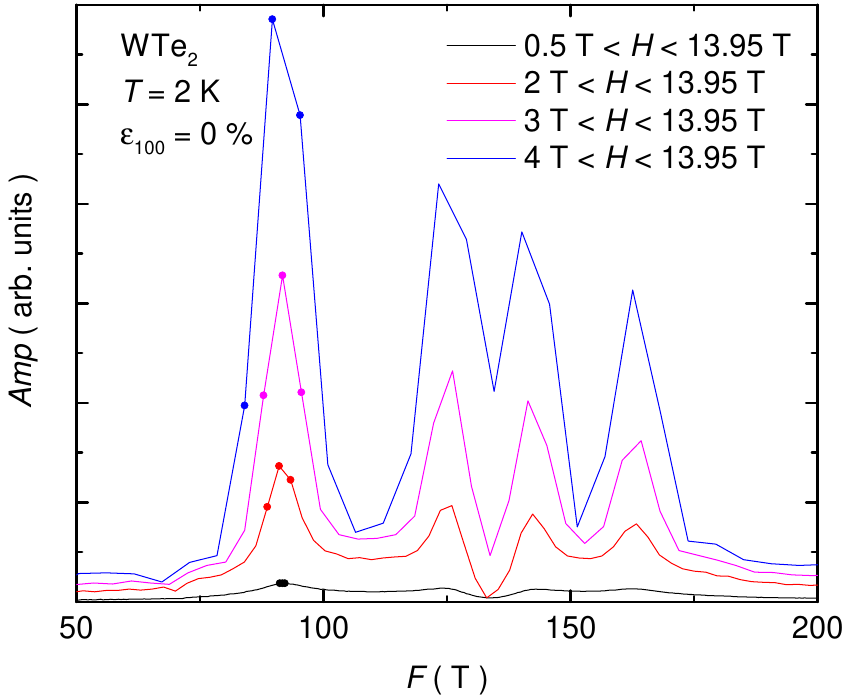}%
	\caption{\textbf{Resolution of frequencies} FFT results of WTe$_{2}$ at $T\,=\,2$\,K and $\epsilon\,=\,0$\,$\%$ for various magnetic field ranges. The filled circles are the data point at $F^{1}$ and the nearest data points around the $F^{1}$.    
		\label{Hrange}}
\end{figure}

When using a FFT to determine the frequencies of the quantum oscillations, the resolution of frequencies is determined by the size of the Fourier window. Figure\,\ref{Hrange} highlights the data spacing around the local maximum which defines $F^{1}$; the data points are more closely spaced for wider field range windows. In order to resolve the strain dependence of frequency changes in WTe$_{2}$, within a relatively small strain range, the magnetic field range of 0.5\,T\,$<\,H\,<$\,13.95\,T was used for all strains.

\begin{figure}
	\includegraphics[scale=1]{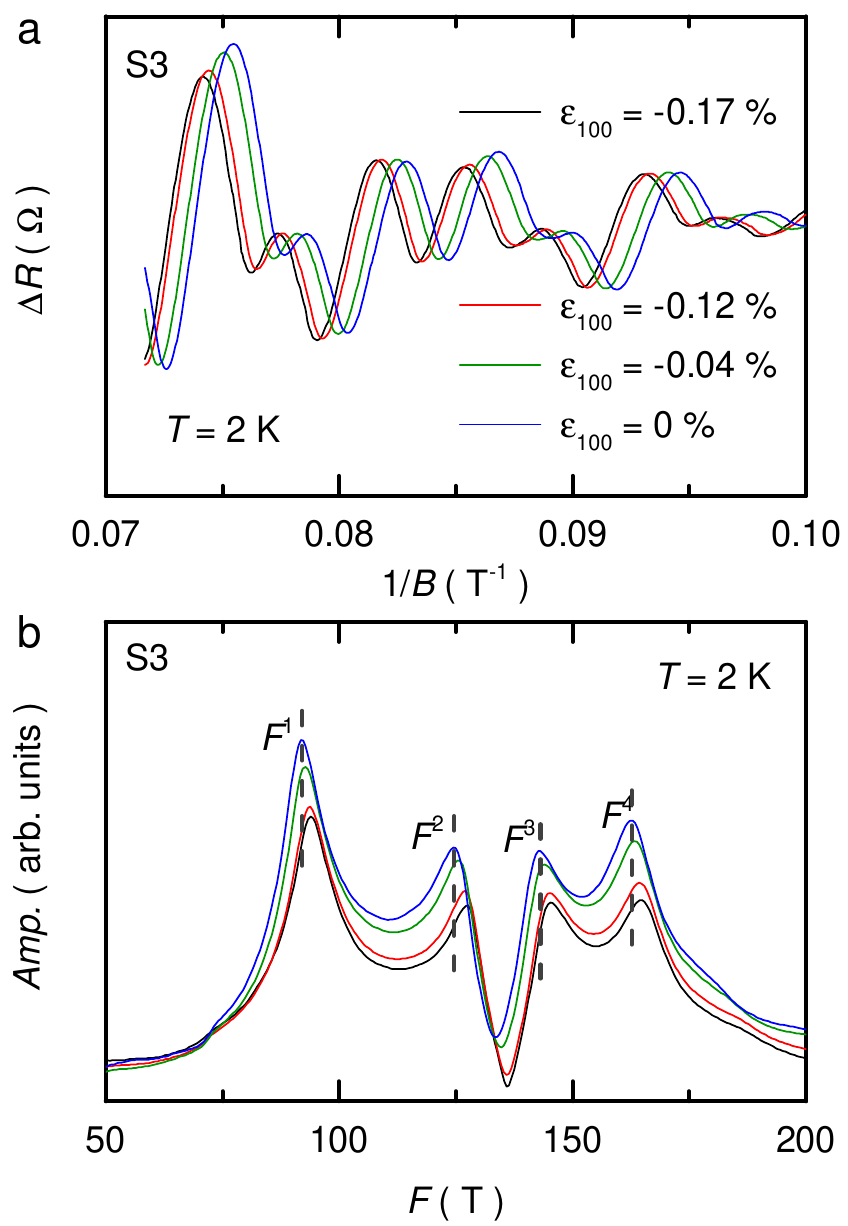}%
	\caption{\textbf{Confirmation of the frequency changes with strain} \textbf{a}, SdH oscillations at various strains for S3. \textbf{b}, FFT results at various strains, and gray dashed lines are guide lines for shifting of the frequencies. 
		\label{doublecheck}}
\end{figure}

In order to check the reproducibility, we performed similar experiments and analysis on S3 with a fewer number of strains. As shown in Fig.\,\ref{doublecheck}a and b, S3 also shows similar behavior as S2 (Fig.\,3). 

Although the resolution is important to resolve the peaks, one also needs to consider the amplitude of peaks to get an effective mass. In this case, it is important to choose the magnetic field range from where the MR starts to show quantum oscillations at the highest temperature. To be more specific, the amplitude of peaks can be reduced due to an artifact of the FFT analysis which comes from the low magnetic field MR data without quantum oscillation. In addition, quantum oscillations at higher temperatures require higher magnetic fields to start. Thus, one needs to determine the magnetic field range based on the highest temperature data that will be used for the analysis. With all these points in mind, we used the magnetic field range of 5\,T\,$<\,H\,<$\,13.95\,T to infer the amplitude changes as a function of the temperature with different strains. 

\subsection{\texorpdfstring{Electrical conductivity from density functional theroy (DFT) calculation}{space}}

\begin{figure}
	\includegraphics[scale=0.8]{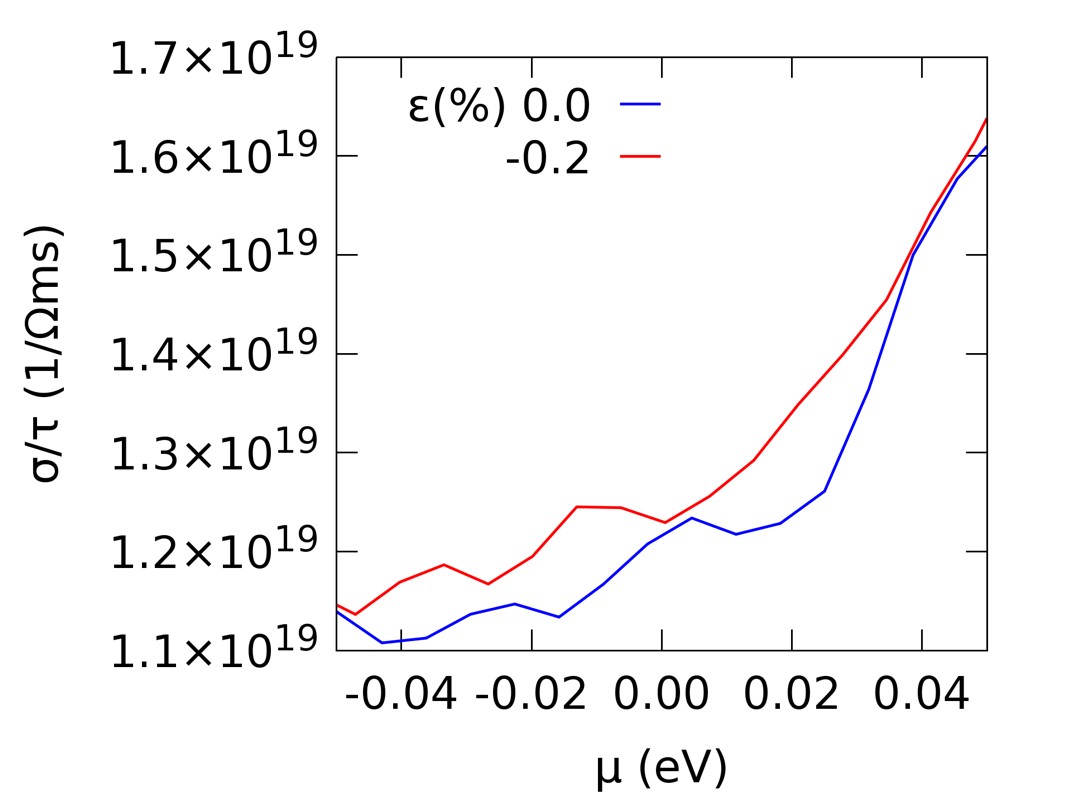}%
	\caption{\textbf{DFT calculation of Conductivity as a function of chemical potential} Conductivity over relaxation time ($\sigma/\tau$) vs. electronic chemical potential ($\mu$) as calculated from semi-classical Boltzmann model with DFT band structure for strain ($\epsilon$) at 0 (blue line) and -0.2\,$\%$ (red line).      
		\label{DFT}}
\end{figure}

The conductivity is calculated from the semi-classical Boltzmann equation with the interpolated DFT band structures.~\cite{Madsen2006} Conductivity over relaxation time ($\sigma/\tau$) without and with ($\epsilon\,=$\,-0.2\,$\%$) are plotted as a function of chemical potential in Fig.\,\ref{DFT}. The result shows positive elastoresistance(It is positive since for negative strain the conductivity goes up) within the chemical potential range of -0.04\,$<\,\mu\,<$\,0.04\,eV. This may be an additional indication that the pure electronic term gives positive elastoresistance in WTe$_{2}$.  

\begin{figure}
	\includegraphics[scale=0.8]{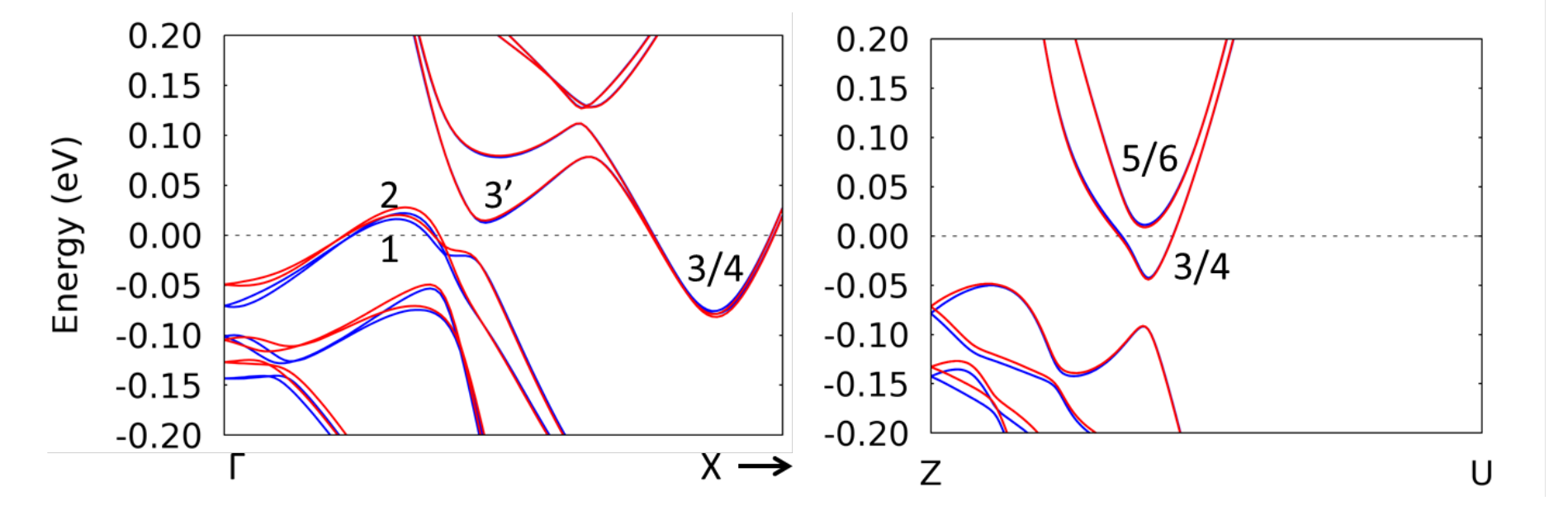}%
	\caption{\textbf{DFT calculation of band structures} DFT band structure for strain ($\epsilon$) at 0 (blue line) and -0.2\,$\%$ (red line) along $\Gamma$-X and Z-U. Band 1 and 2 are the hole bands and band 3 and 4 are the electron bands in Fig.\,3. Another set of electron bands (5 and 6) along Z-U are also close to Fermi level. The label i/j means the and i and j overlap with each other and have the same dispersion in the specific pocket region. The label i'means the second pocket region is also of interest because it is near the Fermi level.    
		\label{DFT_SI}}
\end{figure}

\begin{table}[]
	\caption{Curvature and Edge for each bands in the pocket regions (Fig.\,\ref*{DFT_SI}) that are of interest for the modeling of conductivity}
	\begin{tabular}{cccccc}
		\hline
		Curvature ( eV~A$^{2}$ ) & 1     & 2     & 3'    & 3/4    & 5/6   \\ \hline
		0.0                                   & -18.4 & -19.2 & 26.6  & 35.8   & 15.1  \\
		-0.2                                  & -16.9 & -18.2 & 28.1  & 37.1   & 15.4  \\ \hline
		Edge ( eV )                           & 1     & 2     & 3'    & 3/4    & 5/6   \\ \hline
		0.0                                   & 0.028 & 0.033 & 0.024 & -0.068 & 0.012 \\
		-0.2                                  & 0.032 & 0.039 & 0.026 & -0.071 & 0.010 \\ \hline
	\end{tabular}
\end{table}

\subsection{\texorpdfstring{Modeling and analysis}{space}}

To interpret the (magneto-)elastoresistance (MER) measurements we employ an effective low-energy three-band model. This simplified model can account for the salient features of the (magneto-)elastoresistance (or resistivity rather), in particular its non-monotonic behavior in zero magnetic field and its increase and rapid saturation in finite field. As input it uses the strain-induced changes in the electronic bandstructure, that we infer from experiment and DFT calculations. The minimal model consists of one hole and one electron pocket that cross the Fermi energy $E_F$ at $T=0$ as well as another hole pocket slightly below $E_F$~(see Fig.~4a). This captures the essence of the more complicated bandstructure of WTe$_2$ observed within ARPES~\cite{Pletikosic2014,wu2015,Wu2017} and first-principle calculations~\cite{DiSante2017} (see Fig.~3a) with two pairs of (almost degenerate) electron and two pairs of (almost degenerate) hole pockets, and an additional hole pocket around the $\Gamma$ point with a flat dispersion. (There are additional electron pockets slightly above the Fermi energy, see Fig.~\ref{DFT_SI}. Taking them into account does not change our main qualitative conclusions). The three bands can be characterized by effective masses $m^*_\alpha$ with $\alpha = e, lh, hh$ corresponding to electron ($e$), light-hole $(lh)$ and heavy-hole ($hh$) pockets. We can approximate the effective masses using DFT calculations, neglecting small anisotropies in momentum space. We note that at $T=0$, our simplified model calculation described below agrees with a more detailed DFT transport calculation (see Fig.~\ref{DFT}) that takes the anisotropies in momentum space properly into account. 

Let us first discuss the case of zero magnetic field $\mathbf{B} = 0$. Using a semiclassical Drude-Boltzmann approach and within the quadratic band approximation, one arrives at the well-known Drude formula for the conductivity $\sigma_{\alpha} = n_\alpha e^2/(\Gamma_\alpha m^*_\alpha)$, where $n_\alpha$ is the carrier density and $\Gamma_\alpha$ is the scattering rate of band $\alpha$. The contributions from different bands add in parallel $\sigma = \sum_\alpha \sigma_\alpha$ and the total resistivity is given by $\rho = \sigma^{-1}$. The elastoresistivity is now governed by a sum of different contributions
\begin{align}
\label{eq:1}
\frac{1}{\rho(0)} \frac{d \rho(0)}{d \epsilon} &= \sum_\alpha \frac{\sigma_{\alpha}(0)}{\sigma(0)} \left[\frac{\zeta^{(\alpha)}_m}{m^{*}_\alpha} + \frac{\zeta^{(\alpha)}_\Gamma}{\Gamma_\alpha} - \frac{\zeta^{(\alpha)}_n}{n_\alpha}\right] \,,
\end{align}
where $\rho(0) \equiv \rho(\mathbf{B} = 0)$ and we have introduced the strain derivatives $\zeta^{(\alpha)}_m = \frac{d m^*_\alpha}{d \epsilon}$, $\zeta^{(\alpha)}_\Gamma = \frac{d \Gamma_\alpha}{d \epsilon}$ and $\zeta^{(\alpha)}_n = \frac{d n_\alpha}{d \epsilon}$. Increasing $m^*_\alpha$ and $\Gamma_\alpha$ increases $\rho$, while increasing the carrier density $n_\alpha$ reduces $\rho$. Contributions from different bands are weighted according to their contribution to the total conductivity.

To make progress and estimate $d \Gamma/d \epsilon$, we will use the scattering rates derived within Boltzmann theory and relate $d\Gamma_{\alpha}/d\epsilon$ to changes in the density of states and the phonon properties. The scattering rate consists of a temperature independent impurity part and a $T$-dependent phonon part due to scattering off (mainly acoustic) phonons. One finds~\cite{SmithJensen-Transport-Book} 
\begin{align}
\label{eq:2}
\frac{1}{\Gamma^{(\alpha)}_{\text{imp}}}\frac{d \Gamma^{(\alpha)}_{\text{imp}}}{d \epsilon} &=  \frac{\zeta^{(\alpha)}_m}{m^*_\alpha} + \frac{1}{3} \frac{\zeta^{(\alpha)}_{n}}{n_\alpha}\\
\label{eq:3}
\frac{1}{\Gamma^{(\alpha)}_{\text{ph}}}\frac{d \Gamma^{(\alpha)}_{\text{ph}}}{d \epsilon} &= - \frac{\zeta^{(\alpha)}_m}{m^*_\alpha} - 4 \frac{\zeta_{c}}{c_s}\,,
\end{align}
where $\zeta_c = \frac{d c_s}{d \epsilon}$ is the strain induced change of the phonon velocity. More generally, strain may affect $\Gamma^{(\alpha)}_{\text{ph}}$ in a more complicated way as well, for example, by modifications of the phonon polarization. Under compressive strain, one generally expects that the acoustic phonon velocity increases (hardening) such that $\zeta_c < 0$, due to a change of the Young modulus under strain. As this is a higher order effect in the strain, we expect it to be subleading compared to the change of the carrier densities $\zeta^{(\alpha)}_n$ and the effective masses $\zeta^{(\alpha)}_m$.

Here, we have assumed parabolic and isotropic bands. A more realistic estimate of the strain induced change, in particular of the electron-phonon scattering rate, requires detailed modeling beyond our current work. One can therefore consider $\frac{d \Gamma^{(\alpha)}_{\text{ph}}}{d \epsilon}$ also as a phenomenological parameter of our theory. Using the Matthiessen rule, the total scattering rate is then given by $\Gamma^{(\alpha)} = \Gamma^{(\alpha)}_{\text{imp}} + \Gamma^{(\alpha)}_{\text{ph}}$], which yields the elastoresistitivy in zero field as
\begin{align}
\label{eq:4} 
\frac{1}{\rho(0)} \frac{d \rho(0)}{d \epsilon} &= \sum_{\alpha} \frac{\sigma_\alpha(0)}{\sigma(0)} \left[ \frac{\zeta^{(\alpha)}_m}{m^*_\alpha} \left( 1 + \frac{\Gamma^{(\alpha)}_{\text{imp}} - \Gamma^{(\alpha)}_{\text{ph}}}{\Gamma_\alpha}\right) + \frac{\zeta^{(\alpha)}_{n}}{n_\alpha} \left( \frac{1}{3} \frac{\Gamma^{\alpha}_{\text{imp}}}{\Gamma_\alpha} - 1 \right) - 4 \frac{\zeta_c}{c_s} \frac{\Gamma^{(\alpha)}_{\text{ph}}}{\Gamma_\alpha} \right] \,.
\end{align}
At low temperatures $\Gamma^{(\alpha)}_{\text{ph}} \ll \Gamma^{(\alpha)}_{\text{imp}}$, the elastoresistivity is determined solely by electronic terms and bands that cross the Fermi energy. The behavior of the hole pocket below $E_F$ is thus not relevant to the strain response at low temperature. At low $T$, Eq.~\eqref{eq:4} simplifies to 
\begin{align}
\label{eq:400}
\frac{1}{\rho(0)} \frac{d\rho(0)}{d \epsilon} = 2 \sum_{\alpha} \frac{\sigma_\alpha}{\sigma} \bigl( \frac{\zeta^{(\alpha)}_m}{m^*_\alpha} - \frac{1}{3}\frac{\zeta^{(\alpha)}_{n}}{n_\alpha}\bigr)
\end{align}
From analyzing quantum oscillations, we find that both strain derivatives have the same sign $\zeta^{(\alpha)}_m, \zeta^{(\alpha)}_{n} < 0$ such that the two effects compete with each other. Which of the two dominates depends on microscopic details. In Fig.~1c, we find that the sign of the elastoresistance in the low temperature regime is different for different samples, while the magnitude rapidly decreases at low $T$ (see Fig.~1). We can understand the increase of the carrier density ($\zeta^{(\alpha)}_n < 0$) within our effective three-band model description by noticing that the heavy-hole band is pulled up in energy by strain while the electron band is lowered. As a result, some electrons are being redistributed from the hole to the electron band, incresing the total number of carriers $n = n_e + n_h$ (keeping $\Delta n = n_e - n_h$ fixed). We thus find a positive contribution to the elastoresistivity from $\zeta_n$ within our model (see Fig.~4c at low temperatures).

From the data in Fig.~3e and f, we find that $\zeta^{(\alpha)}_{n}$ and $\zeta^{(\alpha)}_m$ are of the same order of magnitude. It is therefore difficult to estimate which dominates and unambigously predict the sign. In fact, we estimate that the strain-induced enhancement of the mass slightly dominates, which yields $\frac{1}{\rho(0)} \frac{d \rho(0)}{d \epsilon} < 0 $. One should keep in mind, however, that quantum oscillations measure the cyclotron mass, which is a property of an extremal orbit, and not the effective mass $m_i^{-1} = \frac{\partial^2 E_{\boldsymbol{k}}}{\partial k_i \partial k_i}$ (or rather the velocity $v_i = \frac{\partial E_{\boldsymbol{k}}}{\partial k_i}$) in the current direction $i$ (averaged over the Fermi surface), which is the quantity relevant for transport~\cite{SmithJensen-Transport-Book}. In fact, a more detailed microscopic DFT transport calculation predicts a positive slope for $\rho/\tau$, in agreement with experiment (this DFT analysis neglects changes in the scattering time though). 

At higher temperatures, one must take into account not only the strain-induced modifications of the bandstructure right at the chemical potential $\mu$, but also further away from it within a range of $k_B T$. Strain also affects the carriers in thermally (de)populated bands. This is the dominant effect in semiconductors, where strain leads to a redistribution of carriers among valleys with different effective masses, leading to a characteristic $1/T$ behavior of the elastoresistance~\cite{Kanda1982}. We indeed find such a $1/T$ dependence of the elastoresistance at high temperatures $T \gtrsim 250$~K. However, in WTe$_2$ the behavior is much richer due to the presence of both electron and hole carriers. 

The observed increase of the resistivity under compressive strain at intermediate and high $T$ in Fig.~1c, corresponds to $\zeta^{(\alpha)}_{n} > 0$, i.e., a decrease in the carrier density under compressive strain. As shown in Fig.~4c our three-band model calculation reproduces such a behavior (at intermediate temperatures) using the rigid band energy shift trends under strain obtained from DFT as input (see Fig.~3a). The non-monotonic behavior of the elastoresistivity as a function of temperature arises within our model from the fact that the heavy-hole pocket contribetes to transport only at finite $T$, where it is partially filled. Within DFT, strain lifts this pocket in energy by an amount $\Delta D_{\text{hh}}$ that about a factor of 10 times larger than found for the other two pockets. As a result, the dominant effect is a redistribution of holes from the light to the heavy hole pockets, resulting in an increase of $\rho$ at intermediate and larger temperatures. We note that the effective mass enhancement observed within quantum oscillations now adds with the same sign to this term, making the elastoresistance more negative.

Let us now turn to the analysis of the magneto-elastoresistance (MER), i.e., the elastoresistance in finite magnetic field:
\begin{equation}
\text{MER}(T, B) \equiv \frac{1}{R(T,B, \epsilon=0)} \frac{d R(T,B, \epsilon)}{d \epsilon}\biggr|_{\epsilon=0}
\label{eq:M1}
\end{equation}
Experimentally, as shown in Fig.~2, we observe a rich behavior that can be described as follows: at high temperatures, where the magnetoresistance (MR) vanishes, the magneto-elastoresistance (MER) follows the zero-field elastoresistance (ER). At lower $T$, when the (non-saturating) MR first becomes finite and then exceptionally large, we find an increase of the MER proportional to $B^2$ (at small $B$ fields). At temperatures $T \lesssim 50$~K, the observed increase of MER as a function of $B$ crosses over into saturation at a positive plateau value. The MER saturation field strength $B_1(T)$ decreases with temperature $T$. As we show below, the saturation field scale $B_1$ can be associated with the coefficient of the quadratic B-field dependence of the non-saturating MR. At the lowest temperatures $T < 10$~K, where MR exhibits SdH oscillations, the MER exhibits a delicate and strong dependence on the magnetic field, ranging from $\text{MER}(T_0, B) = +120$ to $\text{MER}(T_0, B') =-80$ in the field range of less than one Tesla: $|B-B'| \approx 0.5$~T (see Fig.~2). 

The interesting behavior of MER quantum oscillations in the quantum regime can be straightforwardly understood from the strain-induced change of the SdH oscillation frequency that we have experimentally observed. It occurs due to strain tuning the size of the extremal orbits (see above for more details). We find that the orbits increase under compressive strain, which is in agreement with the prediction of the three-band model that the carrier densities increase at low temperatures. At fixed temperature and field, strain can move the minima and maxima of the oscillating resistance such that a from a position close to the maximum of a SdH oscillation in zero strain becomes a position close to a minimum. This results in a large change of the resistance (see Fig~\ref{QOeffect}). The reverse situation can occur at $B$-field values close by, thus explaining the sign change of $\Delta R_{\epsilon, B}$ when tuning $B$ over a small range.  

\begin{figure}
	\includegraphics[scale=1.5]{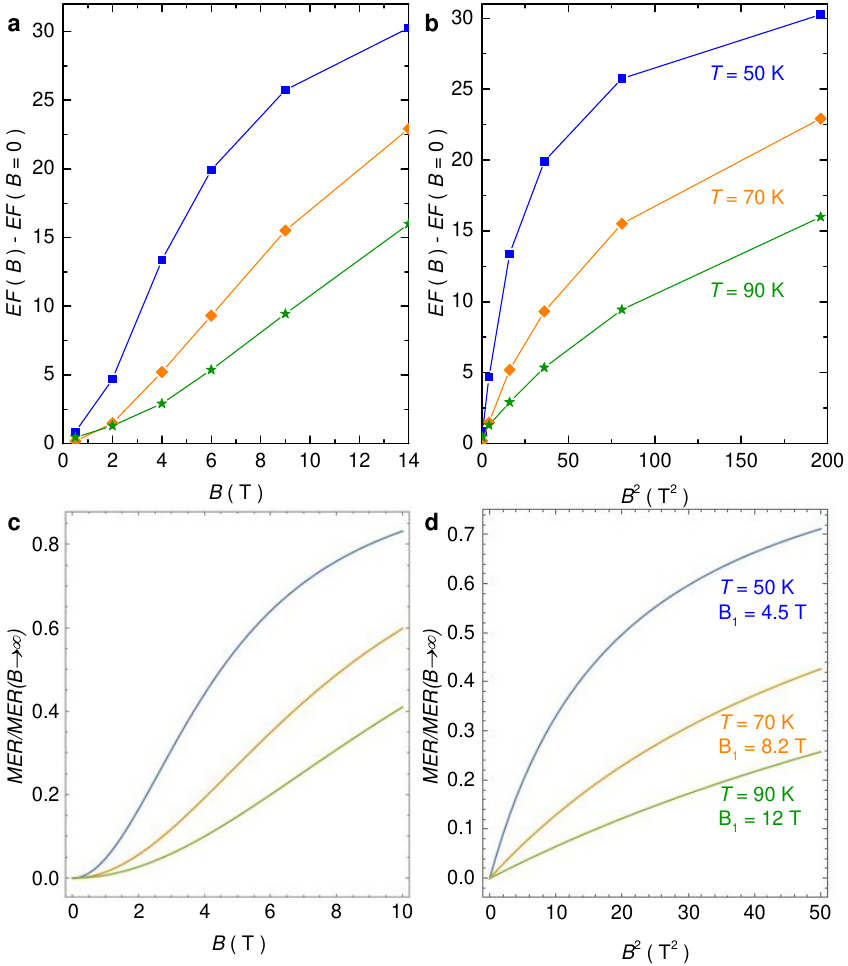}%
	\caption{\textbf{MER data and simulation} \textbf{a}, the magnetic field~($B$) dependence data of $EF\,(B)\,-EF\,(0)$ at 50\,K~(blue), 70\,K~(orange) and 90\,K~(green). \textbf{b}, the $B^{2}$ dependence data of $EF\,(B)\,-EF\,(0)$. \textbf{c}, the magnetic field dependence simulation results of $EF\,(B)\,-EF\,(0)$ at 50\,K~(blue), 70\,K~(orange) and 90\,K~(green). $B_{1}$ value is extracted from Ref.\,\onlinecite{Ali2014}. \textbf{d}, the $B^{2}$ dependence simulation results of $EF\,(B)\,-EF\,(0)$.
		\label{MER}}
\end{figure}

We will now show that the observations in the intermediate temperature regime $10~\text{K} < T < 200~\text{K}$ can be qualitatively captured within a semiclassical two-band model description~\cite{Pippard1989} of electron and hole carriers, where the resistivity takes the well-known form
\begin{align}
\label{eq:6}
\rho(B) &= \rho(0) \left( \frac{1 + e \rho(0) (n_e \mu_h + n_h \mu_e) \mu_e \mu_h B^2}{1 + [e \rho(0) \mu_e \mu_h (n_e - n_h)  B]^2}\right) \,.
\end{align}
The mobilities are given by $\mu_\alpha = e/(m^*_\alpha \Gamma_\alpha)$. It was shown that the carrier compensation in WTe$_2$ is almost perfect and $\Delta n = n_e - n_h$ is very small~\cite{ali2015}. As a result, MR exhibits purely quadratic dependence on magnetic field with no signs of saturation. 
The magnetic field range we consider $B < 14$~T thus lies in the ``intermediate'' regime with $B_1(T) < B < B_{\text{sat}}(T)$ with 
\begin{align}
B_1(T) &= e \rho(0) [(n_e \mu_h + n_h \mu_e) \mu_e \mu_h]^{-1/2} \\
B_{\text{sat}} &= (e \rho(0) \mu_e \mu_h |n_e - n_h|)^{-1} \,.
\label{eq:SI-2}
\end{align}
In this regime of intermediate field strength, MER takes the form
\begin{align}
\label{eq:7}
\frac{1}{\rho(B)} \frac{d \rho(B)}{d \epsilon} - \frac{1}{\rho(0)} \frac{d \rho(0)}{d \epsilon} &= \frac{\rho(0)}{\rho(B)} \frac{B^2}{e^2} \biggl[ \frac{1}{\rho(0)}\frac{d \rho(0)}{d \epsilon} \sum_{\alpha} \Bigl( \frac{\sigma_\alpha}{\sigma(0)} \frac{\sigma^2_{\bar{\alpha}}}{n^2_{\bar{\alpha}}} \Bigr) \nonumber \\ 
& \qquad + \sum_{\alpha} \frac{\sigma_\alpha}{\sigma(0)} \frac{\sigma^2_{\bar{\alpha}}}{n^2_{\bar{\alpha}}} \Bigl( \frac{\zeta^{(\alpha)}_\alpha}{n_\alpha} - \frac{\zeta^{(\alpha)}_m}{m^*_\alpha} - \frac{\zeta^{(\alpha)}_\Gamma}{\Gamma_\alpha} - \frac{2 \zeta^{(\bar{\alpha})}_m}{m^*_{\bar{\alpha}}} - \frac{2 \zeta^{(\bar{\alpha})}_\Gamma}{\Gamma_{\bar{\alpha}}} \Bigr) \biggr] \,,
\end{align}
where $\bar{\alpha}= e (h)$ for $\alpha = h (e)$. We thus predict a deviation of the finite field ER that increases as $B^2$ for small fields and reaches saturation when $B \gg B_1$ (due to the term $\rho(B)$ in the denominator). We can unambigously determine the sign of the term in the bracket, if we approximate $\mu_e \approx \mu_h \equiv \mu$, which seems to be a approximately valid in WTe$_2$~\cite{Luo2015}. Under this assumption, the expression simplifies to 
\begin{align}
\label{eq:8}
\frac{1}{\rho(B)} \frac{d \rho(B)}{d \epsilon} - \frac{1}{\rho(0)} \frac{d \rho(0)}{d \epsilon} &=  2 \mu^2 B^2 \frac{\rho(0)}{\rho(B)} \sum_{\alpha} \frac{\sigma_\alpha}{\sigma(0)} \left( - \frac{\zeta^{(\bar{\alpha})}_m}{m^*_{\bar{\alpha}}} - \frac{\zeta^{(\bar{\alpha})}_{\Gamma}}{\Gamma_{\bar{\alpha}}}\right) \,.
\end{align}
The coefficient of the $B^2$ term is thus positive as $\zeta^{(\alpha)}_m/m^*_\alpha < 0$. Moreover, the saturation value is also positive, in agreement with experiment.(Fig.\,\ref{MER}) Interestingly, the saturation plateau of the MER measures a different combination of strain derivatives as the zero field ER. In particular, it does not explicitly depend on the change of carrier density $\zeta^{(\alpha)}/n_\alpha$ (compare with Eq.~\ref{eq:1}). Combining both measurements thus allows to gain more insight into the electronic response of the material under strain. Let us discuss one particular example. We find that the saturation value of MER at large fields $B \gg B_1$ is positive for all $T$, while ER changes sign as a function of $T$. This contrasting behavior can be traced back to a non-monotonic behavior of the strain-induced change of the carrier densities $\zeta^{(\alpha)}_n$ as a function of $T$ (as opposed to $\zeta^{(\alpha)}_m, \zeta^{(\alpha)}_\Gamma$), as this derivative $\zeta^{(\alpha)}_n$ only occurs on ER (but not in the saturation value of MER).

Finally, we note that at even larger field strengths of $B \gg B_{\text{sat}}$ beyond the regime studied here, the semiclassical analysis predicts yet another combination of strain derivatives $\lim_{B \gg B_{\text{sat}}}\frac{1}{\rho(B)} \frac{d \rho(B)}{d \varepsilon} = \frac{1}{\Delta n^2} \sum_\alpha \frac{n_\alpha}{\sigma_\alpha} \left( \frac{\zeta^{(\alpha)}}{n_\alpha} + \frac{\zeta^{(\alpha)}_m}{m^*_\alpha} + \frac{\zeta^{(\alpha)}_\Gamma}{\Gamma_\alpha} \right)$. Interestingly, the strain derivative of the carrier density $\zeta^{(\alpha)}_n$ now occurs with the opposite sign than at zero field, predicting a sign change if this is the dominant effect (as we believe to be the case in WTe$_2$). Note that we have used that the compensation level $\Delta n$ cannot be tuned by strain due to charge conservation, as long as the quadratic band approximation is valid.

\clearpage

\section*{References}
\bibliographystyle{naturemag}

\begin{thebibliography}{10}
	\expandafter\ifx\csname url\endcsname\relax
	\def\url#1{\texttt{#1}}\fi
	\expandafter\ifx\csname urlprefix\endcsname\relax\def\urlprefix{URL }\fi
	\providecommand{\bibinfo}[2]{#2}
	\providecommand{\eprint}[2][]{\url{#2}}
	
	\bibitem{gere2001}
	\bibinfo{author}{Gere, J.~M.}
	\newblock \emph{\bibinfo{title}{Mechanics of Materials, ; Brooks}}
	(\bibinfo{publisher}{Cole, Pacific Grove, CA}, \bibinfo{year}{2004}).
	
	\bibitem{Kuczynski1954}
	\bibinfo{author}{Kuczynski, G.~C.}
	\newblock \bibinfo{title}{Effect of elastic strain on the electrical resistance
		of metals}.
	\newblock \emph{\bibinfo{journal}{Phys. Rev.}} \textbf{\bibinfo{volume}{94}},
	\bibinfo{pages}{61--64} (\bibinfo{year}{1954}).
	
	\bibitem{Kolobov2016}
	\bibinfo{author}{Kolobov, A.} \& \bibinfo{author}{Tominaga, J.}
	\newblock \emph{\bibinfo{title}{Bulk TMDCs: Review of Structure and Properties.
			In: Two-Dimensional Transition Metal Dichalcogenides.}}, vol.
	\bibinfo{volume}{239} of \emph{\bibinfo{series}{Springer Series in Materials
			Science}} (\bibinfo{publisher}{Springer, Cham}, \bibinfo{year}{2016}).
	
	\bibitem{manzeli2017}
	\bibinfo{author}{Manzeli, S.}, \bibinfo{author}{Ovchinnikov, D.},
	\bibinfo{author}{Pasquier, D.}, \bibinfo{author}{Yazyev, O.~V.} \&
	\bibinfo{author}{Kis, A.}
	\newblock \bibinfo{title}{2d transition metal dichalcogenides}.
	\newblock \emph{\bibinfo{journal}{Nature Reviews Materials}}
	\textbf{\bibinfo{volume}{2}}, \bibinfo{pages}{17033} (\bibinfo{year}{2017}).
	
	\bibitem{Mun2012}
	\bibinfo{author}{Mun, E.} \emph{et~al.}
	\newblock \bibinfo{title}{{Magnetic field effects on transport properties of
			PtSn$_{4}$}}.
	\newblock \emph{\bibinfo{journal}{Phys. Rev. B}} \textbf{\bibinfo{volume}{85}},
	\bibinfo{pages}{035135} (\bibinfo{year}{2012}).
	
	\bibitem{Ali2014}
	\bibinfo{author}{Ali, M.~N.} \emph{et~al.}
	\newblock \bibinfo{title}{{Large, non-saturating magnetoresistance in
			WTe$_{2}$}}.
	\newblock \emph{\bibinfo{journal}{Nature}} \textbf{\bibinfo{volume}{514}},
	\bibinfo{pages}{205--208} (\bibinfo{year}{2014}).
	
	\bibitem{kang2015}
	\bibinfo{author}{Kang, D.} \emph{et~al.}
	\newblock \bibinfo{title}{Superconductivity emerging from a suppressed large
		magnetoresistant state in tungsten ditelluride}.
	\newblock \emph{\bibinfo{journal}{Nature communications}}
	\textbf{\bibinfo{volume}{6}}, \bibinfo{pages}{7804} (\bibinfo{year}{2015}).
	
	\bibitem{Hohenberg1964}
	\bibinfo{author}{Hohenberg, P.} \& \bibinfo{author}{Kohn, W.}
	\newblock \bibinfo{title}{{Inhomogeneous Electron Gas}}.
	\newblock \emph{\bibinfo{journal}{Phys. Rev.}} \textbf{\bibinfo{volume}{136}},
	\bibinfo{pages}{B864--B871} (\bibinfo{year}{1964}).
	
	\bibitem{Kohn1965}
	\bibinfo{author}{Kohn, W.} \& \bibinfo{author}{Sham, L.~J.}
	\newblock \bibinfo{title}{{Self-Consistent Equations Including Exchange and
			Correlation Effects}}.
	\newblock \emph{\bibinfo{journal}{Phys. Rev.}} \textbf{\bibinfo{volume}{140}},
	\bibinfo{pages}{A1133--A1138} (\bibinfo{year}{1965}).
	
	\bibitem{Zeng2015}
	\bibinfo{author}{Zeng, F.}, \bibinfo{author}{Zhang, W.~B.} \&
	\bibinfo{author}{Tang, B.~Y.}
	\newblock \bibinfo{title}{{Electronic structures and elastic properties of
			monolayer and bilayer transition metal dichalcogenides MX$_{2}$ (M = Mo, W; X
			= O, S, Se, Te): A comparative first-principles study.}}
	\newblock \emph{\bibinfo{journal}{Chinese Phys B}}
	\textbf{\bibinfo{volume}{24}}, \bibinfo{pages}{097103}
	(\bibinfo{year}{2015}).
	
	\bibitem{Rourke2012}
	\bibinfo{author}{Rourke, S.~R., P. M. C.;~Julian}.
	\newblock \bibinfo{title}{{Numerical extraction of de Haas-van Alphen
			frequencies from calculated band energies.}}
	\newblock \emph{\bibinfo{journal}{Computer Physics Communications}}
	\textbf{\bibinfo{volume}{183}}, \bibinfo{pages}{324--332}
	(\bibinfo{year}{2012}).
	
	\bibitem{Pippard1989}
	\bibinfo{author}{Pippard, A.}
	\newblock \emph{\bibinfo{title}{Magnetoresistance in Metals}}
	(\bibinfo{publisher}{Cambridge university press}, \bibinfo{year}{1989}).
	
	\bibitem{Pletikosic2014}
	\bibinfo{author}{Pletikosi\ifmmode~\acute{c}\else \'{c}\fi{}, I.},
	\bibinfo{author}{Ali, M.~N.}, \bibinfo{author}{Fedorov, A.~V.},
	\bibinfo{author}{Cava, R.~J.} \& \bibinfo{author}{Valla, T.}
	\newblock \bibinfo{title}{{Electronic Structure Basis for the Extraordinary
			Magnetoresistance in WTe$_{2}$}}.
	\newblock \emph{\bibinfo{journal}{Phys. Rev. Lett.}}
	\textbf{\bibinfo{volume}{113}}, \bibinfo{pages}{216601}
	(\bibinfo{year}{2014}).
	
	\bibitem{wu2015}
	\bibinfo{author}{Wu, Y.} \emph{et~al.}
	\newblock \bibinfo{title}{{Temperature-induced Lifshitz transition in
			WTe$_{2}$}}.
	\newblock \emph{\bibinfo{journal}{Phys. Rev. Lett.}}
	\textbf{\bibinfo{volume}{115}}, \bibinfo{pages}{166602}
	(\bibinfo{year}{2015}).
	
	\bibitem{Wu2017}
	\bibinfo{author}{Wu, Y.} \emph{et~al.}
	\newblock \bibinfo{title}{{Three-dimensionality of the bulk electronic
			structure in WTe$_{2}$}}.
	\newblock \emph{\bibinfo{journal}{Phys. Rev. B}} \textbf{\bibinfo{volume}{95}},
	\bibinfo{pages}{195138} (\bibinfo{year}{2017}).
	
	\bibitem{SmithJensen-Transport-Book}
	\bibinfo{author}{Smith, H.} \& \bibinfo{author}{Jensen, H.~H.}
	\newblock \emph{\bibinfo{title}{Transport Phenomena}}
	(\bibinfo{publisher}{Oxford University Press, Oxford, U.K.},
	\bibinfo{year}{1989}).
	
	\bibitem{Brodsky2017}
	\bibinfo{author}{Brodsky, D.~O.} \emph{et~al.}
	\newblock \bibinfo{title}{{Strain and vector magnetic field tuning of the
			anomalous phase in Sr$_{3}$Ru$_{2}$O$_{7}$}}.
	\newblock \emph{\bibinfo{journal}{Science Advances}}
	\textbf{\bibinfo{volume}{3}} (\bibinfo{year}{2017}).
	
	\bibitem{Lee2016}
	\bibinfo{author}{Lee, J.} \emph{et~al.}
	\newblock \bibinfo{title}{{Single- and few-layer WTe$_{2}$ and their suspended
			nanostructures: Raman signatures and nanomechanical resonances}}.
	\newblock \emph{\bibinfo{journal}{Nanoscale}} \textbf{\bibinfo{volume}{8}},
	\bibinfo{pages}{7854--7860} (\bibinfo{year}{2016}).
	
	\bibitem{Hicks2014}
	\bibinfo{author}{Hicks, C.~W.}, \bibinfo{author}{Barber, M.~E.},
	\bibinfo{author}{Edkins, S.~D.}, \bibinfo{author}{Brodsky, D.~O.} \&
	\bibinfo{author}{Mackenzie, A.~P.}
	\newblock \bibinfo{title}{{Piezoelectric-based apparatus for strain tuning}}.
	\newblock \emph{\bibinfo{journal}{Review of Scientific Instruments}}
	\textbf{\bibinfo{volume}{85}}, \bibinfo{pages}{065003}
	(\bibinfo{year}{2014}).
	
	\bibitem{Ceperley1980}
	\bibinfo{author}{Ceperley, B.~J., D. M.;~Alder}.
	\newblock \bibinfo{title}{{Ground-State of the Electron-Gas by a Stochastic
			Method.}}
	\newblock \emph{\bibinfo{journal}{Phys Rev Lett}}
	\textbf{\bibinfo{volume}{45}}, \bibinfo{pages}{566--569}
	(\bibinfo{year}{1980}).
	
	\bibitem{Perdew1981}
	\bibinfo{author}{Perdew, A., J. P.;~Zunger}.
	\newblock \bibinfo{title}{{Self-Interaction Correction to Density-Functional
			Approximations for Many-Electron Systems.}}
	\newblock \emph{\bibinfo{journal}{Phys Rev B}} \textbf{\bibinfo{volume}{23}},
	\bibinfo{pages}{5048--5079} (\bibinfo{year}{1981}).
	
	\bibitem{Mar1992}
	\bibinfo{author}{Mar, A.}, \bibinfo{author}{Jobic, S.} \&
	\bibinfo{author}{Ibers, J.~A.}
	\newblock \bibinfo{title}{{Metal-metal vs tellurium-tellurium bonding in
			WTe$_{2}$ and its ternary variants TaIrTe$_{4}$ and NbIrTe$_{4}$}}.
	\newblock \emph{\bibinfo{journal}{Journal of the American Chemical Society}}
	\textbf{\bibinfo{volume}{114}}, \bibinfo{pages}{8963--8971}
	(\bibinfo{year}{1992}).
	
	\bibitem{Madsen2006}
	\bibinfo{author}{Madsen, G.~K.} \& \bibinfo{author}{Singh, D.~J.}
	\newblock \bibinfo{title}{{BoltzTraP. A code for calculating band-structure
			dependent quantities}}.
	\newblock \emph{\bibinfo{journal}{Computer Physics Communications}}
	\textbf{\bibinfo{volume}{175}}, \bibinfo{pages}{67 -- 71}
	(\bibinfo{year}{2006}).
	
	\bibitem{Kresse1996}
	\bibinfo{author}{Kresse, G.} \& \bibinfo{author}{Furthm\"uller, J.}
	\newblock \bibinfo{title}{{Efficient iterative schemes for ab initio
			total-energy calculations using a plane-wave basis set}}.
	\newblock \emph{\bibinfo{journal}{Phys. Rev. B}} \textbf{\bibinfo{volume}{54}},
	\bibinfo{pages}{11169--11186} (\bibinfo{year}{1996}).
	
	\bibitem{Bloechl1994}
	\bibinfo{author}{Bl\"ochl, P.~E.}
	\newblock \bibinfo{title}{{Projector augmented-wave method}}.
	\newblock \emph{\bibinfo{journal}{Phys. Rev. B}} \textbf{\bibinfo{volume}{50}},
	\bibinfo{pages}{17953--17979} (\bibinfo{year}{1994}).
	
	\bibitem{Monkhorst1976}
	\bibinfo{author}{Monkhorst, H.~J.} \& \bibinfo{author}{Pack, J.~D.}
	\newblock \bibinfo{title}{{Special points for Brillouin-zone integrations}}.
	\newblock \emph{\bibinfo{journal}{Phys. Rev. B}} \textbf{\bibinfo{volume}{13}},
	\bibinfo{pages}{5188--5192} (\bibinfo{year}{1976}).
	
	\bibitem{Kuo2013}
	\bibinfo{author}{Kuo, H.-H.}, \bibinfo{author}{Shapiro, M.~C.},
	\bibinfo{author}{Riggs, S.~C.} \& \bibinfo{author}{Fisher, I.~R.}
	\newblock \bibinfo{title}{{Measurement of the elastoresistivity coefficients of
			the underdoped iron arsenide Ba(Fe$_{0.975}$Co$_{0.025}$)$_{2}$As$_{2}$}}.
	\newblock \emph{\bibinfo{journal}{Phys. Rev. B}} \textbf{\bibinfo{volume}{88}},
	\bibinfo{pages}{085113} (\bibinfo{year}{2013}).
	
	\bibitem{newnham2005}
	\bibinfo{author}{Newnham, R.~E.}
	\newblock \emph{\bibinfo{title}{Properties of materials: anisotropy, symmetry,
			structure}} (\bibinfo{publisher}{Oxford University Press on Demand},
	\bibinfo{year}{2005}).
	
	\bibitem{DiSante2017}
	\bibinfo{author}{Di~Sante, D.} \emph{et~al.}
	\newblock \bibinfo{title}{{Three-Dimensional Electronic Structure of the
			Type-II Weyl Semimetal WTe$_{2}$}}.
	\newblock \emph{\bibinfo{journal}{Phys. Rev. Lett.}}
	\textbf{\bibinfo{volume}{119}}, \bibinfo{pages}{026403}
	(\bibinfo{year}{2017}).
	
	\bibitem{Kanda1982}
	\bibinfo{author}{{Kanda}, Y.} \& \bibinfo{author}{{Kanda}, Y.}
	\newblock \bibinfo{title}{{A graphical representation of the piezoresistance
			coefficients in silicon}}.
	\newblock \emph{\bibinfo{journal}{IEEE Transactions on Electron Devices}}
	\textbf{\bibinfo{volume}{29}}, \bibinfo{pages}{64--70}
	(\bibinfo{year}{1982}).
	
	\bibitem{ali2015}
	\bibinfo{author}{Ali, M.~N.} \emph{et~al.}
	\newblock \bibinfo{title}{{Correlation of crystal quality and extreme
			magnetoresistance of WTe$_{2}$}}.
	\newblock \emph{\bibinfo{journal}{EPL (Europhysics Letters)}}
	\textbf{\bibinfo{volume}{110}}, \bibinfo{pages}{67002}
	(\bibinfo{year}{2015}).
	
	\bibitem{Luo2015}
	\bibinfo{author}{Luo, Y.} \emph{et~al.}
	\newblock \bibinfo{title}{{Hall effect in the extremely large magnetoresistance
			semimetal WTe$_{2}$}}.
	\newblock \emph{\bibinfo{journal}{Appl. Phys. Lett.}}
	\textbf{\bibinfo{volume}{107}}, \bibinfo{pages}{182411}
	(\bibinfo{year}{2015}).
	
\end{thebibliography}

\end{document}